\documentclass[aps,prd,preprint,a4paper,showpacs,nofootinbib,superscriptaddress]{revtex4-2}
\usepackage{bm}
\usepackage{indentfirst}
\usepackage{amsmath}
\usepackage{graphicx}
\usepackage{amssymb}
\usepackage{subfigure}
\usepackage{amssymb}
\usepackage{hyperref}
\usepackage{epstopdf}
\usepackage{cancel}
\usepackage[section]{placeins}

\usepackage[utf8]{inputenc}
\hypersetup{
    colorlinks=true,
    linkcolor=red,
    citecolor=blue,
}
\usepackage{color}
\usepackage[T1]{fontenc}
\usepackage{txfonts}
\usepackage{orcidlink}
\usepackage[title]{appendix}	

\begin{document}

\title{The underlying black hole phase transitions in an Einstein-Maxwell-dilaton model with a holographic critical point}

\author{Hong Guo}
\email{hong_guo@usp.br}
\affiliation{Escola de Engenharia de Lorena, Universidade de São Paulo, 12602-810, Lorena, SP, Brazil}

\author{Xiao-Mei Kuang}
\email{xmeikuang@yzu.edu.cn}
\affiliation{Center for Gravitation and Cosmology, College of Physical Science and Technology, Yangzhou University, 225009, Yangzhou, China}

\author{Wei-Liang Qian}
\email{wlqian@usp.br (corresponding author)}
\affiliation{Escola de Engenharia de Lorena, Universidade de São Paulo, 12602-810, Lorena, SP, Brazil}
\affiliation{Faculdade de Engenharia de Guaratinguet\'a, Universidade Estadual Paulista, 12516-410, Guaratinguet\'a, SP, Brazil}
\affiliation{Center for Gravitation and Cosmology, College of Physical Science and Technology, Yangzhou University, 225009, Yangzhou, China}

\begin{abstract}
The Einstein-Maxwell-dilaton model exhibits a first-order phase transition curve that terminates at a holographic critical endpoint, offering intriguing insights into the phase diagram of the dual system living on the boundary.
However, the specific instability within the underlying spacetime that triggers the formation of the hairy black hole remains somewhat obscure.
This raises the question of whether one of the hairy black hole phases represents a superconducting or scalarized state and how the two distinct phases merge into a indistinguishable one at the critical point.
This work investigates the associated black hole phase transition and the underlying instabilities by exploring a specific Einstein-Maxwell-dilaton model.
The approach aims to provide transparent insights into the black hole phase transitions in the bulk.
By introducing a nonminimal coupling between a massive real scalar field and a Maxwell field in a five-dimensional anti-de Sitter spacetime, we identify two types of scalarization corresponding to tachyonic instabilities in the ultraviolet and infrared regions.
These distinct instabilities lead to a first-order phase transition between two phases in the $\mu-T$ phase diagram.
Furthermore, this first-order transition terminates at a critical point, beyond which the curve turns back, and the transition becomes a numerically elusive third-order one.
Although one does not encounter a critical endpoint, the model still offers a consistent interpretation for the observed {\it cross-over} in the low baryon density region.
We analyze the thermodynamic properties of the scalarized hairy black holes and discuss the implications of our findings.
\end{abstract}

\maketitle

\newpage

\tableofcontents

\section{Introduction}\label{sec=intro}

Phase transitions of the strongly interacting system under extreme conditions governed by quantum chromodynamics (QCD) play a crucial role in high-energy nuclear collisions and in the evolution of the early universe. 
Research in this area has been the main focus of the Relativistic Heavy Ion Collider (RHIC)~\cite{Harrison:2003sb,2014APS..DNP.FJ001C} and the Large Hadron Collider (LHC)~\cite{Morrissey:2009tf, Wells:2008xg} experiments. 
On the one hand, non-perturbative lattice QCD calculations indicate that, for small chemical potentials, there is a smooth cross-over between the chiral and deconfinement/confinement phase transitions~\cite{Borsanyi:2010cj, Borsanyi:2013bia, HotQCD:2014kol}. 
On the other hand, effective theories such as the Nambu-Jona-Lasinio (NJL) model~\cite{Asakawa:1989bq,Schwarz:1999dj}, the Dyson-Schwinger equation~\cite{Qin:2010nq,Shi:2014zpa} and the functional renormalization
group~\cite{Fu:2019hdw, Fu:2021oaw}, suggest that the system features a first-order phase transition as the chemical potential increases.
Therefore, it has been speculated that the first-order phase transition curve terminates at a critical endpoint (CEP).

An alternative for the non-perturbative QCD endeavor is the holographic approach~\cite{adscft-qcd-03, adscft-qcd-05, adscft-qcd-07, adscft-qcd-08, adscft-qcd-emd-10, Kim:2012ey, Cai:2012xh, DeWolfe:2010he, adscft-qcd-40, Erdmenger:2020flu}. 
Through the gauge/gravity duality, the thermodynamic properties and characteristic evolution of black holes in a weakly coupled gravitational system are mapped to phase transitions in a strongly coupled system described by the field theory living on the anti-de Sitter (AdS) boundary. 
In terms of the Einstein-Maxwell-dilaton (EMD) model~\cite{DeWolfe:2010he, DeWolfe:2011ts}, holographic theories have played a significant role in studying topics related to the QCD matter~\cite{adscft-qcd-emd-11, adscft-qcd-emd-12, Chen:2022goa, Fu:2024wkn, Li:2024lrh}. 
Recently, improved holographic models~\cite{Critelli:2017oub, Grefa:2021qvt}, by matching lattice QCD data, have achieved a better quantitative description of hot QCD with 2+1 flavors and physical quark masses. 
These models have obtained satisfactory results in studies of the expected QCD phase diagram and the existence of CEP~\cite{Knaute:2017opk, Cai:2022omk, Chen:2024ckb, Liu:2023pbt}.
The EMD models typically consist of a real scalar field, referred to as dilaton, and a U(1) gauge field.
On the one hand, the condensation of the scalar field breaks the conformal invariance and, therefore, mimics the running coupling of QCD.
On the other hand, the Abelian gauge field is dual to the baryon number current, which furnishes a nonvanishing chemical potential by turning on an appropriate electric field in the black hole spacetime.

In the context of black hole perturbation theory, tachyonic instability merges in the asymptotically AdS spacetime, complemented by the presence of a Maxwell field and a charged scalar.
Here, the asymptotic AdS spacetime is understood to play a crucial role~\cite{Sudarsky:2002mk} in evading the prerequisite of no-hair theorems primarily formulated in asymptotically flat spacetimes~\cite{Ruffini:1971bza, Chrusciel:2012jk, Carter:1971zc}.
A well-known consequence of such instability is that it gives birth to the celebrated holographic superconductor~\cite{Gubser:2008px, Hartnoll:2008kx}.
An alternative mechanism for hairy black holes, referred to as spontaneous scalarization, has also garnered much attention in recent years.
As was first introduced within the framework of the Einstein-Scalar-Gauss-Bonnet (ESGB) theory~\cite{Antoniou:2017acq, Doneva:2017bvd, Silva:2017uqg}, it is characterized by the nonminimal coupling of a scalar field with the Gauss-Bonnet curvature. 
As the strength of the coupling increases and exceeds a critical value, it eventually introduces a negative effective mass and triggers a tachyonic instability~\cite{Blazquez-Salcedo:2018jnn}.
In particular, such an instability can be established in asymptotically flat spacetimes.
Besides, its emergence can be inferred via the analysis of the quasinormal frequencies of the underlying ``bald'' black hole solutions~\cite{Doneva:2017bvd, Blazquez-Salcedo:2018jnn, Myung:2018iyq}, and specifically, the onset of spontaneous scalarization is identified as when the purely imaginary quasinormal modes touch the origin~\cite{Cardoso:2013fwa}. 
Visually, a sufficient condition for such a tachyonic instability is when the effective potential forms a potential well~\cite{buell1995potentials, Cardoso:2013opa}. 
It can be understood as the {\it dynamic} counterpart of the superluminal propagation in the system with a substantial Gauss-Bonnet term~\cite{Brigante:2008gz, Brigante:2007nu, Buchel:2009tt, Hofman:2008ar, Hofman:2009ug, Camanho:2009vw}. 

Besides the higher curvature, it was observed that nonlinear coupling to the electromagnetic field~\cite{Herdeiro:2018wub, Fernandes:2019rez, Zou:2019bpt, Guo:2021zed, Deshpande:2024itz} may also trigger spontaneous scalarization.
Such models are referred to in the literature as the Einstein-Maxwell-Scalar (EMS) theory, where a charged black hole becomes unstable and subsequently acquires a scalar hair.
It is intriguing to note that the relevant degrees of freedom of EMS and EMD theories closely resemble each other.
The EMD model was proposed earlier and has since been employed primarily to analyze the QCD phase diagram.
To our knowledge, the underlying instabilities and the resulting hairy black holes associated with the underlying QCD phase diagram have not been explicitly elaborated.
In this regard, some questions remain unsettled about whether one of the QCD phases represents a superconducting or scalarized black hole. If so, how do the two distinct phases merge into an indistinguishable one at the critical point?

The present study is motivated to gain further insights regarding the underlying instabilities, the properties of the transitions between the resulting hairy black holes, and the overall phase structure.
To this end, we construct a specific EMD model featuring a nonminimal coupling between a massive real scalar field and a Maxwell field in a five-dimensional AdS spacetime.
By construction, the model furnishes a transparent interpretation of the underlying hairy black holes.
In place of a superconducting black hole, the gravitational theory accommodates the two types of scalarized black holes corresponding to two tachyonic instabilities in the ultraviolet and infrared regions.
The first type of scalarization can be turned on by tuning the scalar mass in terms of a mass parameter.
The effective potential of the scalar field is found to become negative at spatial infinity, leading to a tachyonic instability associated with the ultraviolet region~\cite{Hertog:2004bb, Perivolaropoulos:2020uqy, Myung:2018iyq}.
The second type of scalarization arises from the nonminimal coupling with the Maxwell field. 
Its emergence can be triggered by increasing the coupling constant, and the resulting effective potential might form a valley near the horizon, leading to instability in the infrared region discussed in Refs.~\cite{Silva:2017uqg, Myung:2018iyq,Guo:2020sdu, Guo:2024vhq}.
By properly tuning the scalar field mass and the coupling constant, one can assess the corresponding hairy black holes and investigate the phase transition between them. 

The two distinct types of hairy black holes imply a transition between the two phases in the $\mu-T$ phase diagram.
Subsequently, it gives rise to a phase structure that is somewhat different from what one encounters where scalarized and superconducting phases coexist~\cite{Guo:2024vhq}.
Specifically, a first-order transition curve is observed that terminates at a critical point, beyond which the curve turns back, and the phase transition becomes a third-order one.
In other words, one encounters a more smooth phase transition beyond the could-have-been CEP.
It is argued that the model offers a consistent interpretation for the observed cross-over in the low baryon density region.
We analyze the order of the phase transitions based on the Ehrenfest classification and discuss the implications of our findings.

The remainder of the paper is organized as follows. 
In the following section, we elaborate on the proposed Einstein-Maxwell-dilaton model and derive the equations of motion with the corresponding boundary conditions.
In Sec.~\ref{sec=phase}, we investigate the two hairy solutions and the consequential phase diagram, and in particular, the thermodynamic properties near the critical point.
We relegate further analysis of the effective potential to Appx.~\ref{sec=poten}, which is used to explore the instabilities justifying the corresponding hairy black hole solutions.
The last section is devoted to further discussions and concluding remarks.

\section{An alternative Einstein-Maxwell-dilaton model}\label{sec=model}

In this section, we introduce a specific EMD model consisting of a massive real scalar field $\psi$ nonminimally coupled to Maxwell's field $F_{\mu\nu}$ in a five-dimensional AdS spacetime.
The action reads
\begin{equation}\label{eq=action}
    S=\frac{1}{2\kappa_5^2}\int d^5x\sqrt{-g}\Bigg[R-\frac{1}{2}\nabla^\mu\psi\nabla_\mu\psi-\frac{1}{4}f(\psi)F^2_{\mu\nu}-V(\psi)\Bigg],
\end{equation}
where $\kappa^2_5=8\pi G_5$ is Newton's constant in five-dimensional spacetime. 
The scalar field $\psi$ is coupled to the Maxwell background by the nonminimal coupling function $f(\psi)$. 
As a standard choice~\cite{Herdeiro:2018wub} regarding spontaneous scalarization, the form of the coupling function is given by 
\begin{equation}\label{eq=coupling}
    f(\psi)=e^{-\lambda_f \psi^2}
\end{equation}
where $\lambda_f$ denotes the Maxwell coupling parameter. 
The potential is defined by~\cite{DeWolfe:2010he} 
\begin{equation}\label{eq=potentialfunc}
    V(\psi)=-12\cosh{(\lambda_V\psi)},
\end{equation}
where the mass parameter $\lambda_V$ is directly related to the effective scalar mass at spatial infinity as discussed below Eq.~\eqref{eq=inf_matter}.
As elaborated below, the choice of the coupling function~\eqref{eq=coupling} and the potential function~\eqref{eq=potentialfunc} is primarily motivated to host two distinct mechanisms for the black hole solutions in the bulk.
It is noted that different choices of the coupling function might also lead to spontaneous scalarization and have been explored in Ref.~\cite{Fernandes:2019rez}.

We consider a static charged black hole solution that exhibits spatial isotropy and translational invariance, which satisfies the following {\it ansatz}~\cite{Critelli:2017oub}
\begin{align}
    ds^2 &= e^{2A(r)}\big[-h(r)dt^2+d\vec{x}^2\big]+\frac{1}{h(r)}dr^2,\label{eq=metric}\\
    \psi &= \psi(r), \\
    A &=A_\mu dx^\mu=\phi(r)dt.
\end{align}
The radial location of the event horizon is defined by $h(r_h)=0$, and the asymptotic $AdS_5$ boundary is given at the spatial infinity $r\rightarrow\infty$.
The temperature and entropy density in the boundary field theory are holographically associated with the Hawking temperature and the area of the black hole horizon, respectively,
\begin{align}
    T=\frac{e^{A(0)}h'(0)}{4\pi}, \quad s=\frac{2\pi e^{3A(0)}}{\kappa^2_5}.
\end{align}

The equations of motion can be derived by varying the action~\eqref{eq=action} with respect to the matter and metric fields:
\begin{align}
& \psi^{\prime \prime}(r)+\left[\frac{h^{\prime}(r)}{h(r)}+4 A^{\prime}(r)\right] \psi^{\prime}(r)-\frac{1}{h(r)}\left[\frac{\partial V(\psi)}{\partial \psi}-\frac{e^{-2A(r)} \phi^{\prime}(r)^2}{2} \frac{\partial f(\psi)}{\partial \psi}\right]=0, \label{eq=scamotion}\\
& \phi^{\prime \prime}(r)+\left[2 A^{\prime}(r)+\frac{d[\ln (f(\psi))]}{d \psi} \psi^{\prime}(r)\right] \phi^{\prime}(r)=0, \label{eq=maxmotion}\\
& A^{\prime \prime}(r)+\frac{\psi^{\prime}(r)^2}{6}=0, \label{eq=metrmotion1}\\
& h^{\prime \prime}(r)+ 4 A^{\prime}(r) h^{\prime}(r)-e^{-2 A(r)} f(\psi) \phi^{\prime}(r)^2=0, \label{eq=metrmotion}\\
& h(r)\left[24 A^{\prime}(r)^2-\psi^{\prime}(r)^2\right]+6 A^{\prime}(r) h^{\prime}(r)+2 V(\phi) +e^{-2 A(r)} f(\psi) \phi^{\prime}(r)^2=0.\label{eq=consmotion}
\end{align}
Notably, the last line, Eq.~\eqref{eq=consmotion}, is a constraint that can also be derived by combining the independent components of Einstein's equation. 
In addition, the Gauss charge $Q_G$ and the Noether charge $Q_N$ are derived by the first integral of the respective equations of motion, Eqs.~\eqref{eq=maxmotion} and~\eqref{eq=metrmotion}:
\begin{align}
    & Q_G(r)=f(\psi) e^{2 A(r)} \phi^{\prime}(r), \\
    & Q_N(r)=e^{2 A(r)}\left[e^{2 A(r)} h^{\prime}(r)-f(\psi) \phi(r) \phi^{\prime}(r)\right].
\end{align}

To proceed, let us discuss the boundary conditions at the event horizon and spatial infinity. 
We note that the radial location of the event horizon $r_h$ is a free parameter in the metric ansatz Eq.~\eqref{eq=metric}. 
In other words, we can always set $r_h=0$ by rescaling the radial coordinate.
Such a choice also simplifies the numerical calculations by avoiding irrelevant singular points of the differential equations.\footnote{
It is worth noting that different black hole phases are often characterized by their horizon sizes. 
The aforementioned choice shifts this information to a specific scale of the radial coordinate carried by $A(r)$.
As detailed in Sec.~\ref{sec=phase} and Appx.~\ref{sec=poten}, the hairy black hole phases are distinguished by the distinct roles of the potential and coupling functions, as well as their different manifestations in the effective potential.}

Near the event horizon, the regular condition requires that the Maxwell field satisfies $\phi(0)=0$, while $\psi(0) \ne 0$ is due to the nonvanishing scalar hair. 
Additionally, $h(0)=0$ ensures a simple zero for the metric function at the event horizon, $h'(0)=1$ can also be fixed by rescaling the time coordinate, and $A(0)=0$ can be attained by rescaling the space and time coordinates simultaneously by a same scale~\cite{Critelli:2017oub}. 

At spatial infinity, the ultraviolet behaviors of the fields read~\cite{DeWolfe:2010he}
\begin{align}
 A(r) &=\alpha(r)+\mathcal{O}\left(e^{-2 \Delta\alpha(r)}\right), \\ 
 h(r) &=h_0+\mathcal{O}\left(e^{-4 \alpha(r)}\right), \label{eq=inf_metric}\\
 \psi(r) &=\psi_0 e^{-\Delta\alpha(r)}+\mathcal{O}\left(e^{-2 \Delta\alpha(r)}\right), \label{eq=inf_scalar}\\
 \phi(r) &=\phi_0+\phi_2 e^{-2 \alpha(r)}+\mathcal{O}\left(e^{-(2+\Delta) \alpha(r)}\right),\label{eq=inf_matter}
\end{align}
where $\alpha(r)=A_0+A_1 r$. 
As $\psi\rightarrow 0$ near the boundary, we have $V(\psi)=-12-6\lambda_V^2\psi^2+\mathcal{O}(\psi^4)$.
This implies that the cosmological constant is $\Lambda=-6$ and the AdS radius $L=1$.
The asymptotic mass of the scalar field near the boundary is given by $m^2=-12\lambda_V^2$.
It is worth noting that the potential function Eq.~\eqref{eq=potentialfunc} can be rescaled by adjusting the AdS radius, which effectively introduces a new degree of freedom~\cite{Kubiznak:2016qmn}.
Including this degree of freedom may enrich the underlying phase structure.
According to the AdS/CFT correspondence, the asymptotic form of the scalar field derives the scaling dimension of the gauge theory operator as $\Delta_{\pm}=\frac{d\pm\sqrt{d^2+4m^2}}{2}$. 
Due to exponential decay, the suppression of the term related to $\Delta_+$ at infinity is more significant than that of $\Delta_-$. 
One may effectively only retain $\Delta_-$ as the source and rewrite $\Delta_-=\Delta$. 
As posited by the holographic dictionary, the baryon chemical potential and charge density are obtained from the boundary value of the gauge field
\begin{align}
    \mu=\lim_{r\rightarrow\infty}\phi(r)=\phi_0,\quad \rho=\lim_{r\rightarrow\infty}\frac{Q_G}{\kappa^2_5}=-\frac{\phi_2}{\kappa^2_5}.
\end{align}

It is worth noting that the above ansatz and related boundary conditions are mathematically convenient for handling the numerical solutions of differential equation systems. 
Following~\cite{DeWolfe:2010he, Critelli:2017oub}, however, in order to assess physically relevant quantities using holographic duality, it is necessary to rescale the current coordinates into a {\it standard} coordinate system. 
Based on such a rescaling, the thermodynamic quantities can be rewritten as
\begin{align}\label{eq=thermoquan}
& T =\frac{1}{4\pi\psi_0^{1/\Delta}\sqrt{h_0}}, \quad
s =\frac{2 \pi}{\kappa_5^2 \psi_0^{3 / \Delta}}, \quad
\mu =\frac{\phi_0}{\psi_0^{1 / \Delta} \sqrt{h_0}}, \quad
\rho =-\frac{\phi_2}{\kappa_5^2 \psi_0^{3 / \Delta} \sqrt{h_0}}.
\end{align}

It is noted that the above system admits a ``bald'' charged black hole with trivial scalar hair in an asymptotically AdS spacetime, which is essentially a five-dimensional RN-AdS black hole spacetime, with metric and the electromagnetic field possessing the forms
\begin{align}
    &ds^2=-g(r)dt^2+\frac{1}{g(r)}dr^2+r^2d\vec{x}^2,\\
    &g(r)=\frac{r^2}{L^2}-\frac{2M}{r^2}+\frac{Q^2}{r^4},\\
    &\phi(r)=\frac{Q}{r_h^2}\left(1-\frac{r_h^2}{r^2}\right),
\end{align}
where $M$, $Q$, $L$ and $r_h$ are the mass, charge, AdS radius and event horizon of the RN-AdS black hole.
Before delving into an analysis of the specific hairy black holes and the subsequent phase transitions, it is worthwhile to first examine the properties of the effective potential of the perturbed scalar field around this ``bald'' black hole in the probe limit.
Such an analysis has been recently carried out in the context of an ESGB theory~\cite{Guo:2024vhq}, which turned out to accommodate two hairy black holes: a superconducting and a scalarized one.
Besides, a first-order phase transition is observed between the two phases. 
Specifically, the transition curve spans the entire phase space without concluding at any critical point.
This behavior is attributed to the distinct physical origins of the two hairy black holes.  
However, as discussed earlier, in the EMD model, the first-order transition curve is known to terminate at a critical point.
Therefore, it is instructive to explore how the stability of the underlying metric can be triggered by specific causes to gain insight into the differing outcomes between the two scenarios. 
To this end, we consider the static scalar perturbations $\delta\psi$ which satisfies the following Schr\"odinger-type equation in the linear order
\begin{equation}
\frac{d^2 \varphi(r)}{d r_{*}^2}-V_\mathrm{eff}(r) \varphi(r)=0 ,
\end{equation}
where one has introduced $\delta\psi=\frac{\varphi}{r}$ and the tortoise coordinate $r_*=\int\frac{dr}{g(r)}$, and the effective potential is given by
\begin{equation}\label{eq=potential}
    V_\mathrm{eff}(r)=g(r)\left(\frac{3g(r)}{4r^2}+\frac{3g^{\prime}(r)}{2r}+V^0_{\psi}-\frac{f^0_{\psi}}{2}\phi^{\prime}(r)^2 \right),
\end{equation}
where $V^0_\psi=-12\lambda_V$ and $f^0_\psi=-2\lambda_f$ which are essentially the first-order derivatives of the potential and coupling functions of the action~\eqref{eq=action}.
Specifically, we have 
\begin{align}
    \frac{\partial V(\psi)}{\partial \psi} &=V^0_\psi\psi+\mathcal{O}(\psi^3) ,\\
    \frac{\partial f(\psi)}{\partial \psi} &=f^0_\psi\psi+\mathcal{O}(\psi^3) .
\end{align}

The analysis of the effective potential in the Appx.~\ref{sec=poten} indicates two types of tachyonic instabilities, which are controlled by the scalar mass $\lambda_V$ in the ultraviolet region and the strength of the nonminimal coupling $\lambda_f$ in the infrared region, respectively.
Subsequently, they could trigger two different hairy black holes. 
Furthermore, an intriguing finding is that these two types of instability can be controlled solely by the electric charge of the black hole $Q$.
The explicit competition between the two instabilities suggests a first-order phase transition between the two underlying hairy black holes, which involves until the two hairy black hole states merge continuously and become indistinguishable, giving rise to  a critical point.
Specifically, the two-phase equilibrium is governed by the Gibbs condition, while this study adopts the Ehrenfest classification of phase transition, for which the order is governed by the lowest-order derivative of the free energy that exhibits a discontinuity.
Notably, this phase diagram is drastically different when compared to the ESGB model~\cite{Guo:2024vhq}, where a first-order phase transition curve traverses the entire phase space, owing to the distinct nature of the two phases.
In the following section, we numerically construct the two hairy black holes and show the presence of the critical point.

\section{Hairy black hole solutions and phase structure }\label{sec=phase}

In this section, we first numerically solve the equations of motion for hairy black hole solutions under different model parameters.
Subsequently, we explore the structure of the phase diagram by exploring the Gibbs condition between the two hairy black hole states.
In particular, based on the Ehrenfest classification, we assess the order of the phase transition.
By explicitly evaluating the derivatives of the free energy density using the AdS/CFT dictionary, we observe a third-order phase transitin beyond the critical point, in the place of a CEP.

Based on the preceding analysis of the effective potential, we demonstrate that the system exhibits two types of hairy black holes. 
For simplicity, in what follows, we will refer to the scalarized solutions triggered by tachyonic instability near the horizon as infrared hairy black holes and those due to instability at large radial coordinates as ultraviolet hairy black holes.

\subsection{Numerical procedure for hairy black holes}\label{sec=phase_sol}

The numerical solutions for hairy black holes are derived by applying the boundary conditions discussed in Sec.~\ref{sec=model}. 
Specifically, one solves the field equations by numerically integrating from the horizon to spatial infinity.
The initial values for the numerical integrations are constrained by performing Taylor expansions of the relevant fields near the horizon, substituting them into the corresponding differential equations, and demanding that the resultant expressions vanish at each order.
Specifically, one results in two independent variables, $\psi_h \equiv \psi(0)$ and $\phi_h \equiv \phi'(0)$, where $\psi_h$ measures the scalar hair condensation near the horizon, while $\phi_h$ is related to the Gauss charge $Q_G$ of the black hole.
In practice, to avoid singularity on the horizon, the initial position of the numerical integration is slightly shifted to $r_\mathrm{start}=10^{-8}$.
The asymptotical behaviors of the fields are then evaluated at $r_\mathrm{end}=10$, where the coefficients defined in Eqs.~(\ref{eq=inf_metric}-\ref{eq=inf_matter}) are numerically extracted and used to calculate the thermodynamic quantities given by Eq.~\eqref{eq=thermoquan}.
As will become clear later, varying the two variables ($\psi_h$ and $\phi_h$) effectively enumerates the two-dimensional phase space expressed in temperature and chemical potential.

\begin{figure}[thbp]
    \centering
    \includegraphics[width=0.48\linewidth]{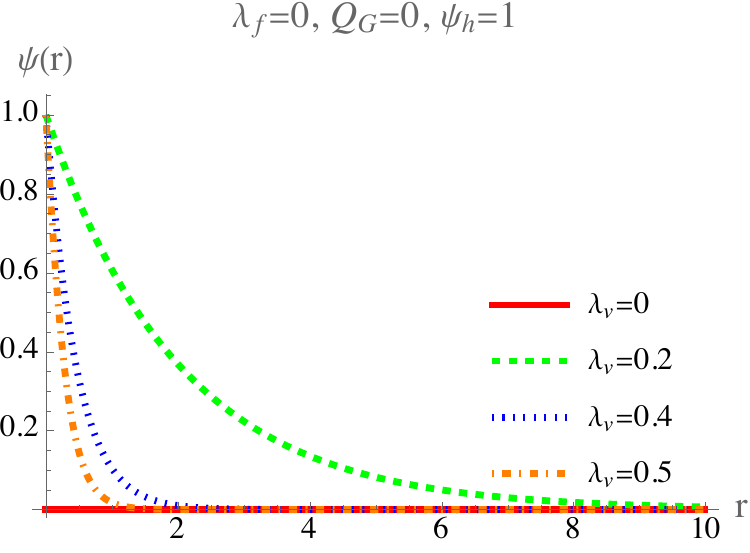}
    \includegraphics[width=0.48\linewidth]{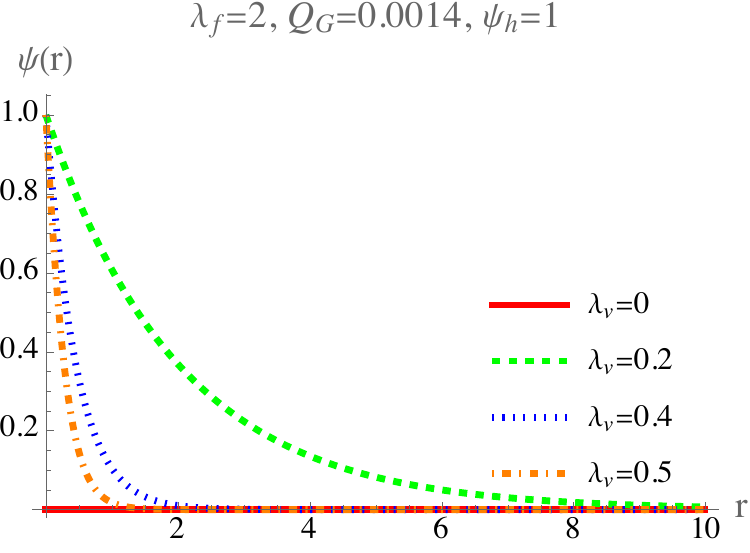}
    \caption{The profiles of the scalar field for the ultraviolet hairy black holes.
    Left: The resulting scalar field with vanishing Maxwell field, evaluated for different mass parameters. 
    Right: The resulting scalar field with a nonvanishing Maxwell field, evaluated for different mass parameters.
    The scalar hair vanishes in both cases with a vanishing mass parameter $\lambda_V$.}
    \label{fig=tahairBH}
\end{figure}

We first explore the profiles of the scalar hair regarding both types of black holes. Branches of solution exist in which the scalar field has no node or has nodes, but the latter are always physically disfavored \cite{Gubser:2008zu} because the spatial oscillations in $\psi(r)$ shall increase the energy density. So we focus on the hairy solutions where $\psi(r)$ has no node.
In Fig.~\ref{fig=tahairBH}, we present the spatial distributions of the scalar field for the ultraviolet hairy black holes.
As shown in the left panel of Fig.~\ref{fig=tahairBH}, the emergence of such a scalar hair does not depend on the Maxwell field, as observed in Refs.~\cite{Hertog:2004dr, Torii:2001pg,Hertog:2004bb}.
On the right panel, we show the corresponding profiles by turning on an insignificant amount of the Maxwell field by assigning a small charge $Q_G\ll 1$ to the black hole.
By comparing the two panels, we note that the profiles of the scalar field are mainly similar, which indicates that the Maxwell field does not play a significant role in this type of scalarization.
Specifically, the scalar field decreases monotonically as the radial coordinate increases and vanishes at spatial infinity.
As a result, the first-order derivative of the scalar field remains negative.
We note that this feature of the ultraviolet hairy black holes persists for vanishing or small black hole charges.
Also, one observes that the strength of the scalar field decreases with decreasing mass parameter and, in particular, the scalar hair vanishes as $\lambda_V\to 0$ for both cases.

\begin{figure}[thbp]
    \centering
    \includegraphics[width=0.48\linewidth]{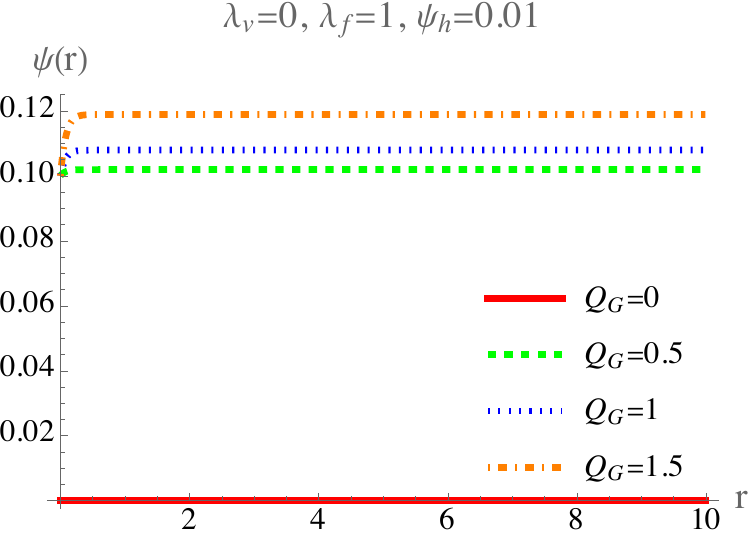}
    \includegraphics[width=0.48\linewidth]{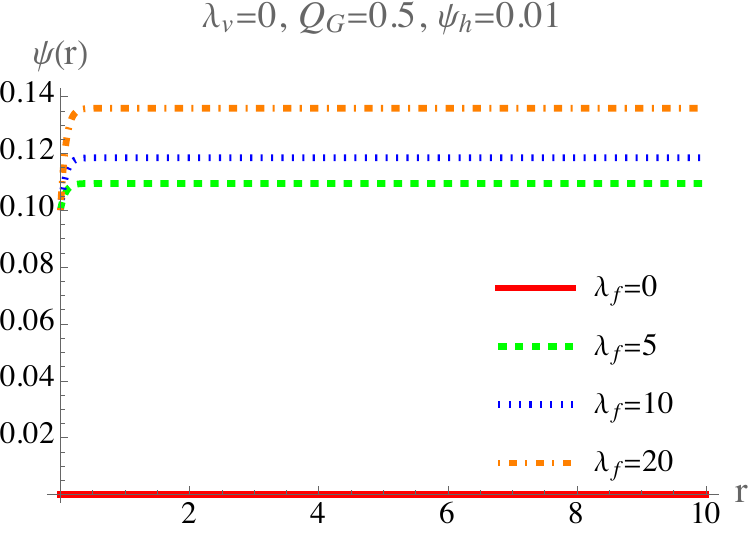}
    \caption{The scalar field profile for the infrared hairy black holes. 
    Left: The resulting scalar field with vanishing mass parameter, evaluated for different black hole charges. 
    Right: The resulting scalar field with vanishing mass parameter, evaluated for different mass parameters.
    The scalar field vanishes when the nonlinear coupling or the black hole charge vanishes.}
    \label{fig=scahairBH1}
\end{figure}

As shown in Fig.~\ref{fig=scahairBH1}, the infrared hairy black holes are readily encountered even when the mass parameter vanishes. 
As discussed above, the asymptotical AdS spacetime does not play a significant role in this case, as the tachyonic instability essentially resides in the infrared region. 
If the nonlinear coupling or the black hole charge vanishes, such a scalar hair will not be formed.
As illustrated in the left panel of Fig.~\ref{fig=scahairBH1}, when the nonlinear coupling $\lambda_f\ne 0$ increases, a scalar field is formed as the black hole charge increases.
It features a rapid increase near the horizon and remains constant at larger radial coordinates.
Specifically, the first-order derivative of the scalar field with respect to the radial coordinate is positive near the horizon and approaches zero as the radial coordinate increases. 
By comparing Figs.~\ref{fig=tahairBH} and~\ref{fig=scahairBH1}, it is observed that the overall features of the scalar hair are rather distinct.
As the mass parameter vanishes $\lambda_V\rightarrow 0$, we have $\Delta=\Delta_-\to 0$, and subsequently, the scalar field does not vanish on the boundary as $\psi(r)=\psi_0 e^{-\Delta\alpha(r)}\to \psi_0$.
Such an asymptotical behavior\footnote{It is noted that this extreme case does not affect the phase diagram elaborated in the next section because the boundary condition Eq.~\eqref{eq=inf_scalar} remains valid as we assume $\lambda_V=0.5$ in the calculations.} is manifestly shown in Fig.~\ref{fig=scahairBH1}.
The strength of the scalar field becomes more significant as the black hole charge increases.
In the right panel, we show the profile's dependence on the nonlinear coupling for a given charge $Q_G$.
The main characteristics of the profile remain unchanged.
It is observed that the scalarization becomes more significant as the coupling's strength increases.

\subsection{Phase transition and QCD phase diagram}\label{sec=thermodynamic}

In a general choice of metric parameters, it is expected that the two types of hairy black holes might co-exist. 
Furthermore, the present subsection investigates the phase transition between the two hairy black holes and the subsequent phase diagram.
To identify the transition curve using the Gibbs condition, we evaluate the thermodynamic quantities utilizing the AdS/CFT dictionary.
The phase transition is, therefore, determined by equating the chemical potential and temperature of the two hairy black hole phases.
In what follows, we enumerate $(\psi_h,\phi_h)$ to effectively scan the two-dimensional phase space in chemical potential and temperature $(\mu, T)$ while fixing the metric parameters $\lambda_V=0.5$ and $\lambda_f=1$.
Specifically, the scalar hair spans the range $\psi_h\in [0.1,4.5]$, while $\phi_h$ is confined by the condition $\phi_h<\phi_h^\mathrm{max}\equiv\sqrt{-2V(\psi_h)/f(\psi_h)}$~\cite{Critelli:2017oub}. 
In practice, regarding the effective potential, one observes that the phase transition emerges in the vicinity where a local potential well is critically formed near the horizon, as shown by the green dashed curve in Fig.~\ref{fig=poten_Q}.
This corresponds to a scenario where the infrared hairy black holes become marginally stable.
Therefore, a numerically convenient approach to roughly estimate the transition curve in the phase space is to identify a particular value $\phi_h^0$ when the first-order derivative of the scalar field near the horizon marginally attains zero from a previously negative value for a given $\psi_h$. 
By and large, for $\phi_h<\phi_h^0$, the ultraviolet hairy black hole solutions are obtained, while for $\phi_h>\phi_h^0$, the relevant black hole solutions are due to tachyonic instability in the infrared region. 
As further explored below, the transition between the two phases is very smooth, demonstrated by an almost continuous change in the profiles of the metric functions and Maxwell field.

\begin{figure}[thbp]
    \centering
    \includegraphics[width=0.68\linewidth]{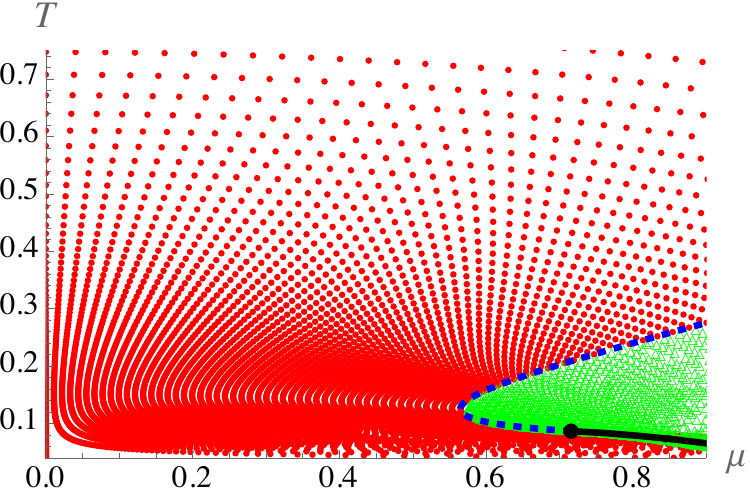}
    \caption{
    The phase structure of the present model shown in the chemical potential vs. temperature plane.
    The solid red dots represent the ultraviolet hairy black hole phase, while the empty green triangles occupy the region associated with the infrared hairy black holes. 
    The solid black curve represents a first-order transition line that terminates at a critical point indicated by a filled black circle, beyond which the blue dashed curve denotes a third-order phase transition line between the two phases.
    The critical point $(\mu_c, T_c)=(0.716, 0.087)$ is indicated by the black dot.}
    \label{fig=phasediag}
\end{figure}

\begin{figure}[thbp]
    \centering
    \includegraphics[width=0.325\linewidth]{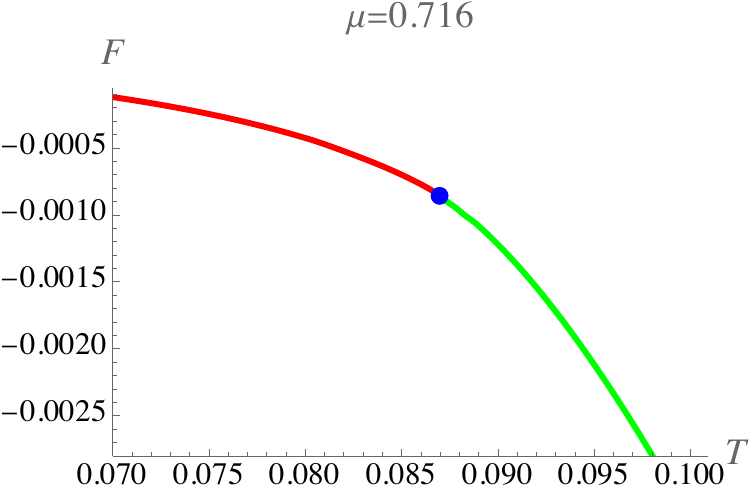}
    \includegraphics[width=0.325\linewidth]{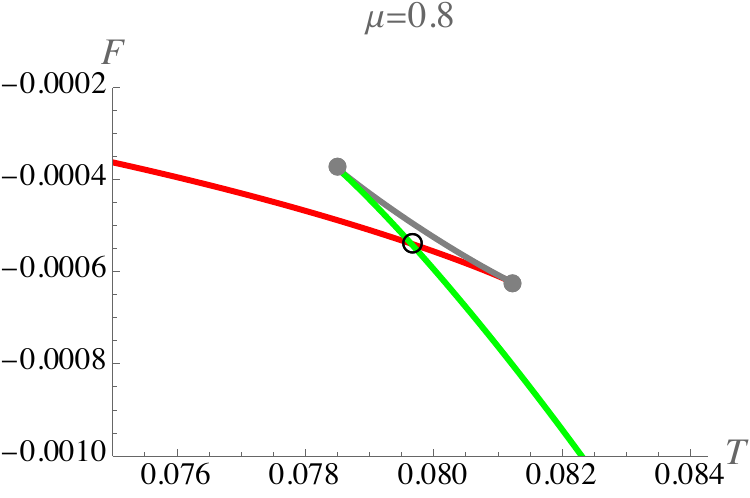}
    \includegraphics[width=0.325\linewidth]{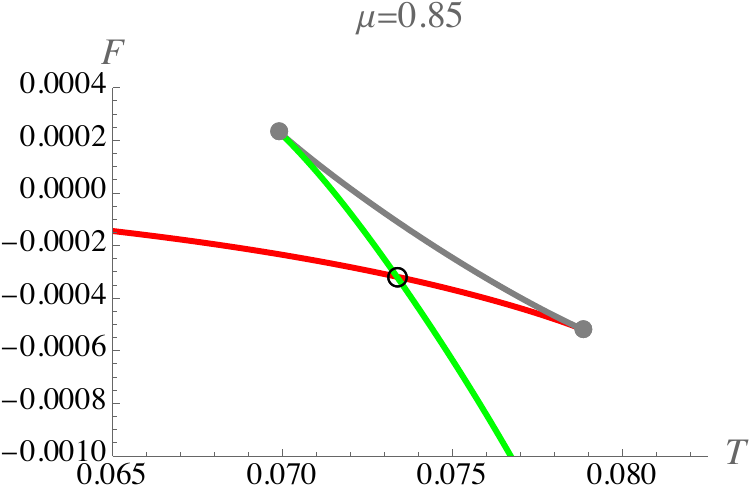}
    \vfill\vspace{10pt}
    \includegraphics[width=0.325\linewidth]{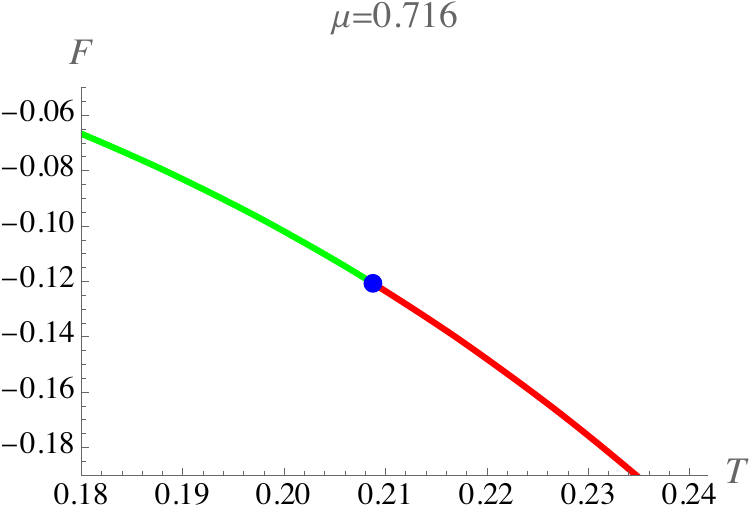}
    \includegraphics[width=0.325\linewidth]{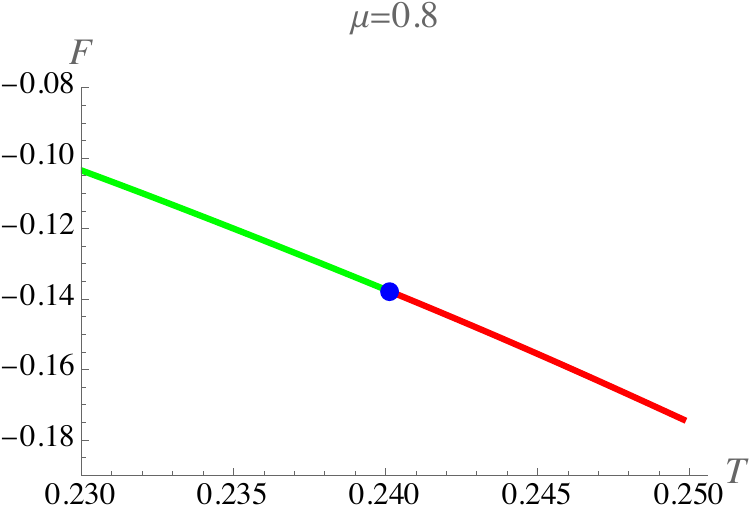}
    \includegraphics[width=0.325\linewidth]{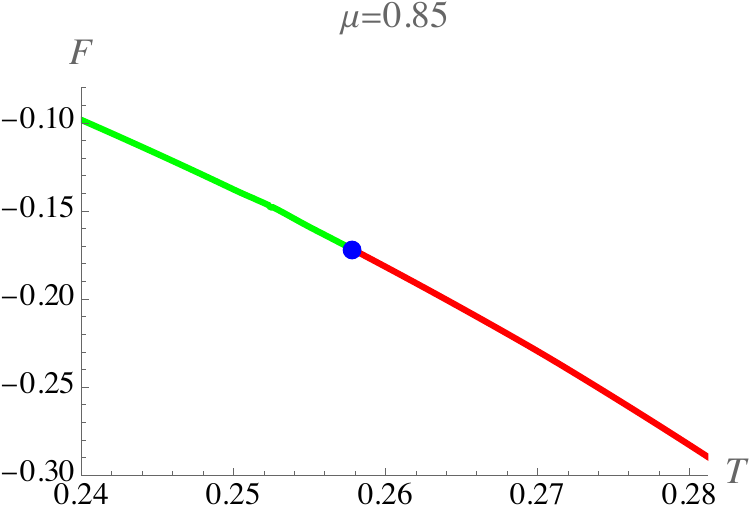}
    \caption{
    The Gibbs free energy density as a function of temperature for different chemical potentials.
    In the first row, the top-middle and top-right plots correspond to two trajectories in the phase space that traverse the first-order transition curve, whereas the top-left plot corresponds to the case where the trajectory goes through the critical point.
    The solid black circle indicates the point of phase transition. 
    Beyond this point, the solid green and solid red curves correspond to the subcooled and superheated states that are metastable.
    The solid gray curve represents hairy black hole solutions that are thermodynamically unstable.
    The second row illustrates three trajectories that traverse the third-order transition curve.
    The solid red curve represents states associated with the ultraviolet hairy black holes, while the solid green curve indicates those related to the infrared hairy black holes.
    }
    \label{fig=freeenergy}
\end{figure}

\begin{figure}[thbp]
    \centering
    \includegraphics[width=0.325\linewidth]{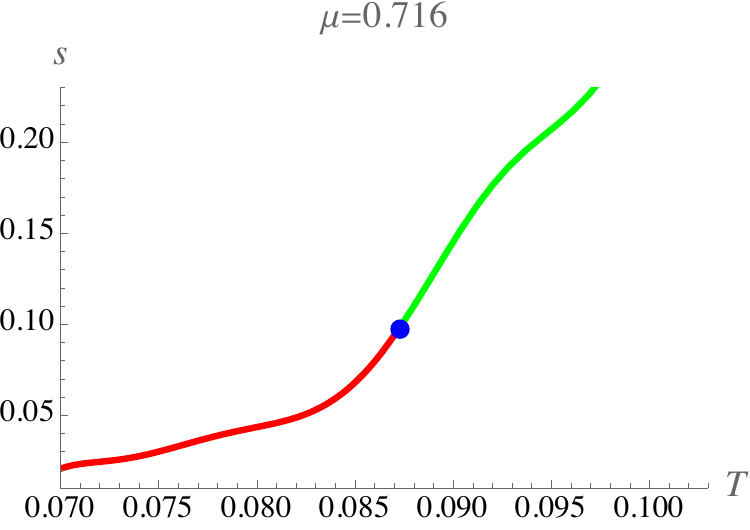}
    \includegraphics[width=0.325\linewidth]{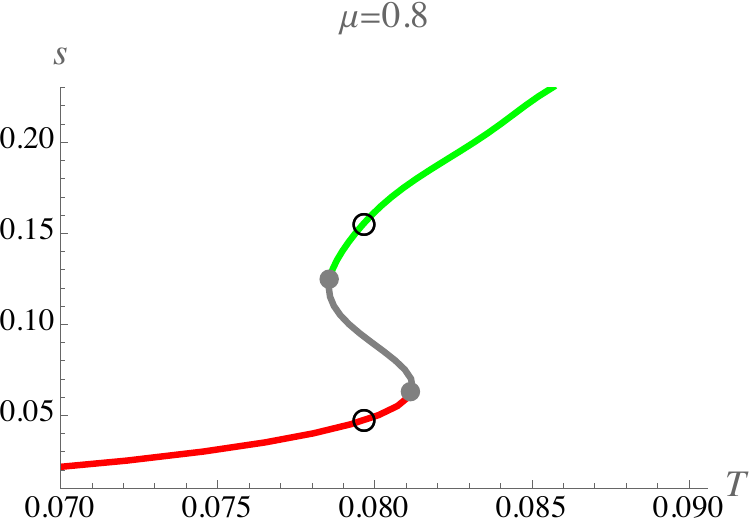}
    \includegraphics[width=0.325\linewidth]{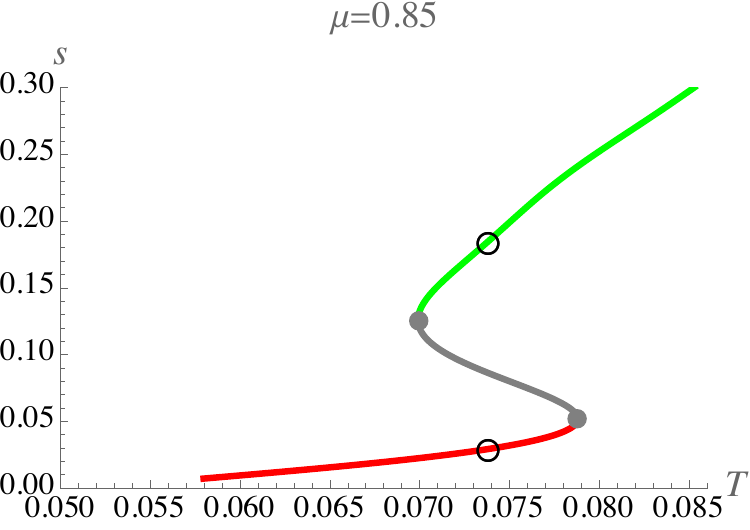}
    \vfill\vspace{10pt}
    \includegraphics[width=0.325\linewidth]{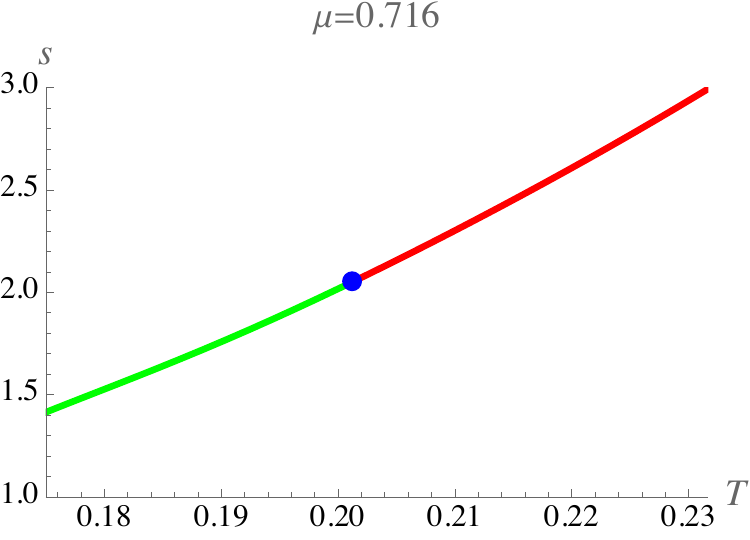}
    \includegraphics[width=0.325\linewidth]{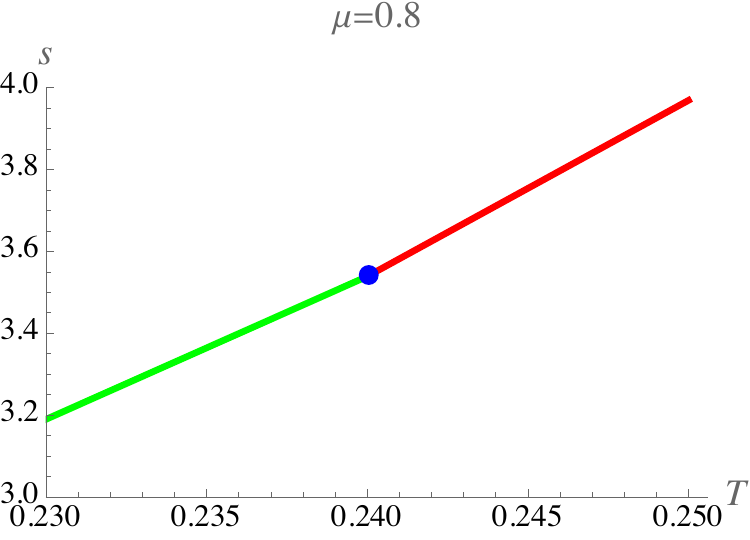}
    \includegraphics[width=0.325\linewidth]{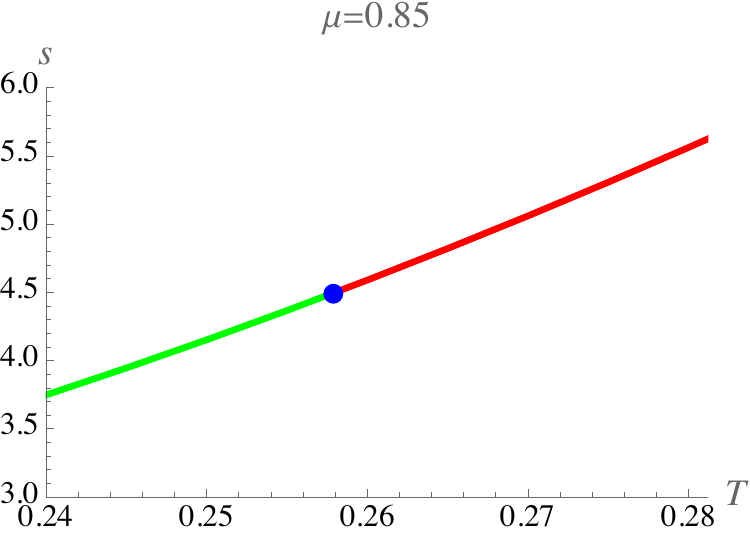}
    \caption{
    The same as Fig.~\ref{fig=freeenergy} but for the entropy density as a function of temperature for different chemical potentials.
    In the first row, the top-middle and top-right plots correspond to first-order transitions, where the entropy density features a continuous ``S'' shape, leading to a finite vertical jump at the transition temperature between the two states indicated by empty black circles. 
    For the top-left plot and those in the second row, the evolutions of the entropy density are continuous, indicating that the transition is beyond the first order.}
    \label{fig=entropy-T}
\end{figure}

\begin{figure}[thbp]
    \centering
    \includegraphics[width=0.325\linewidth]{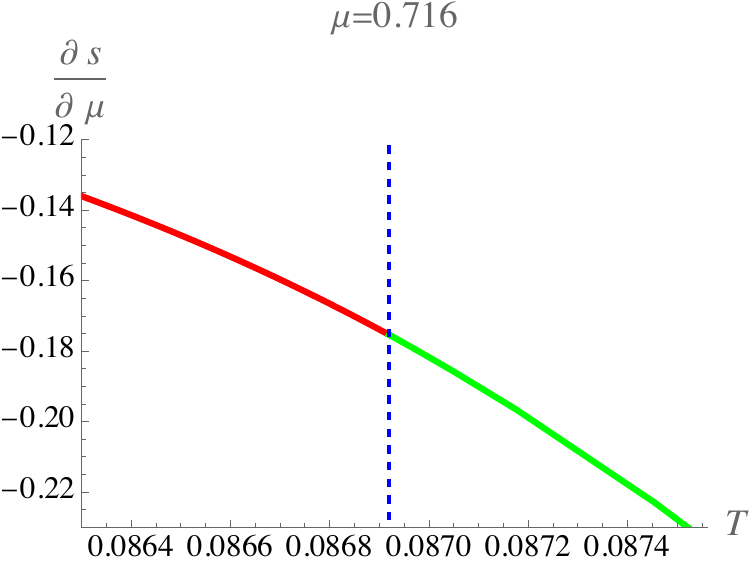}
    \includegraphics[width=0.325\linewidth]{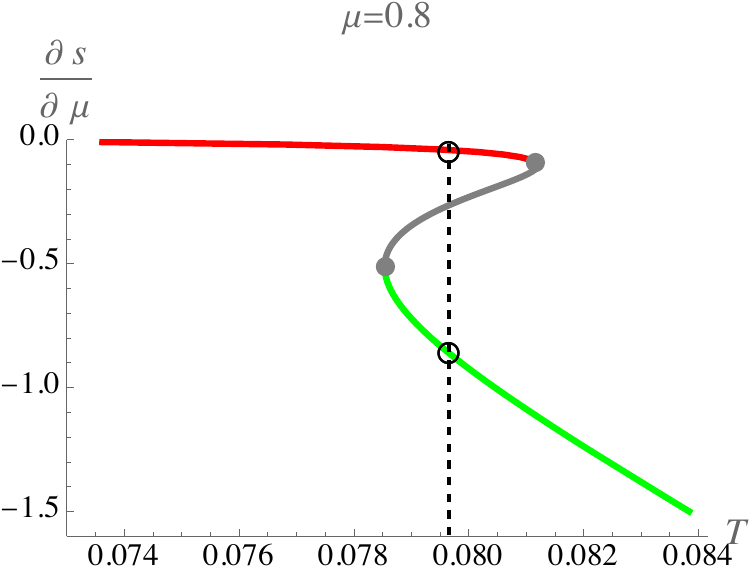}
    \includegraphics[width=0.325\linewidth]{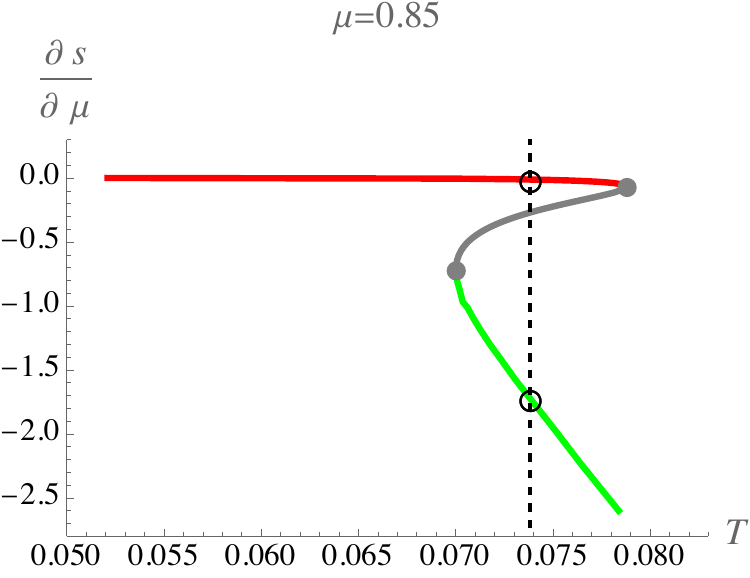}
    \vfill\vspace{10pt} 
    \includegraphics[width=0.325\linewidth]{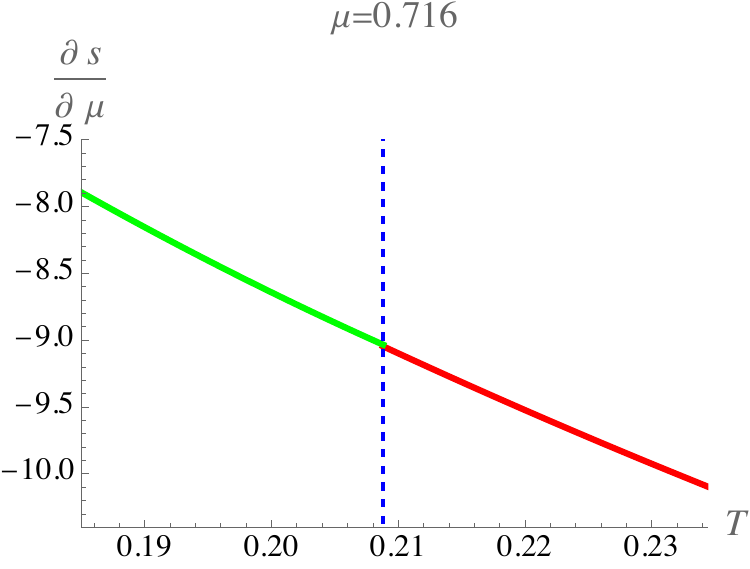}
    \includegraphics[width=0.325\linewidth]{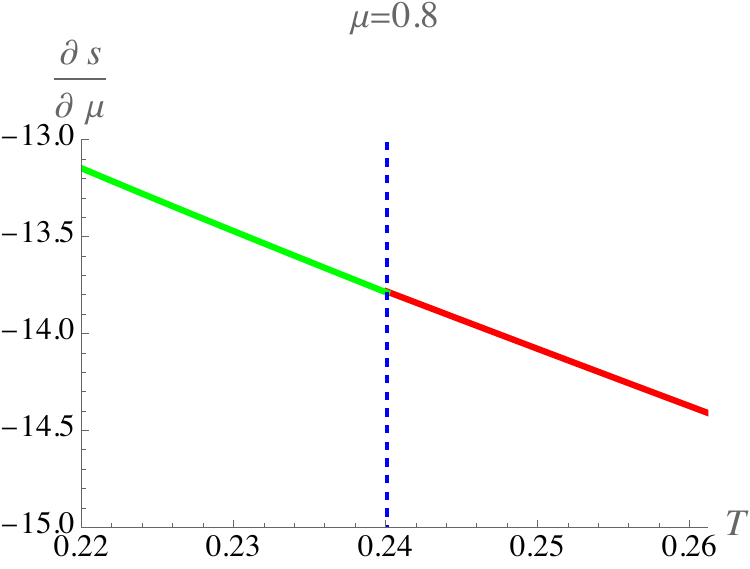}
    \includegraphics[width=0.325\linewidth]{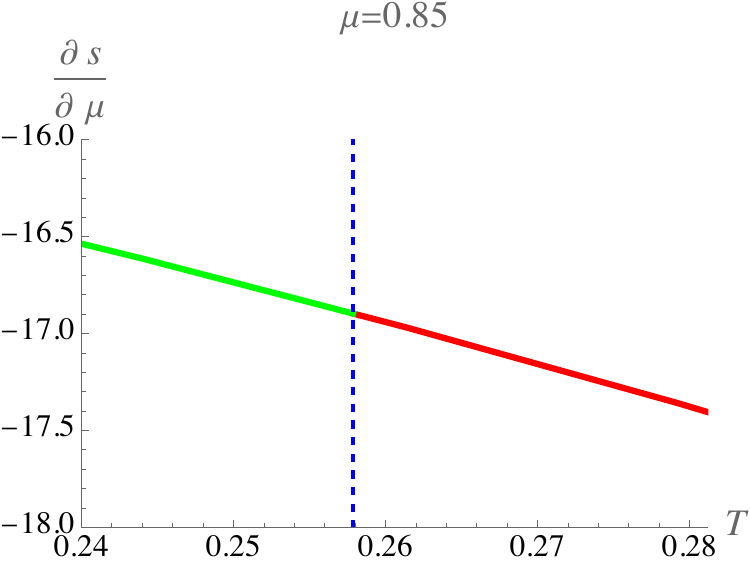}
    \caption{
    The same as Fig.~\ref{fig=freeenergy} but for the first-order derivative of the entropy density as a function of temperature for different chemical potentials.
    In the first row, the top-middle and top-right plots correspond to first-order transitions, where the derivative of the entropy density features a continuous reversed ``S'' shape, leading to a finite vertical jump at the transition temperature between the two states indicated by empty black circles. 
    For the top-left plot and those in the second row, the evolutions of the entropy density are continuous, indicating that the transition is beyond the second order.
    }
    \label{fig=dentropy-T}
\end{figure}

\begin{figure}[thbp]
    \centering
    \includegraphics[width=0.325\linewidth]{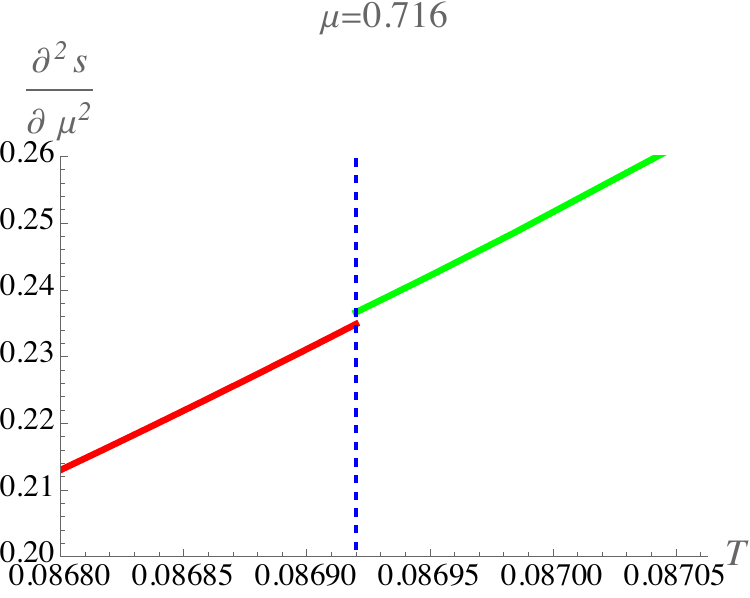}
    \includegraphics[width=0.325\linewidth]{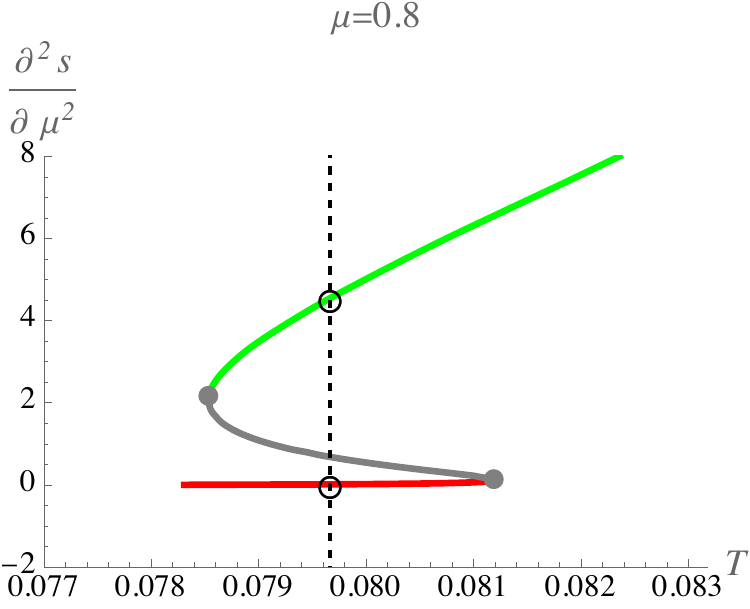}
    \includegraphics[width=0.325\linewidth]{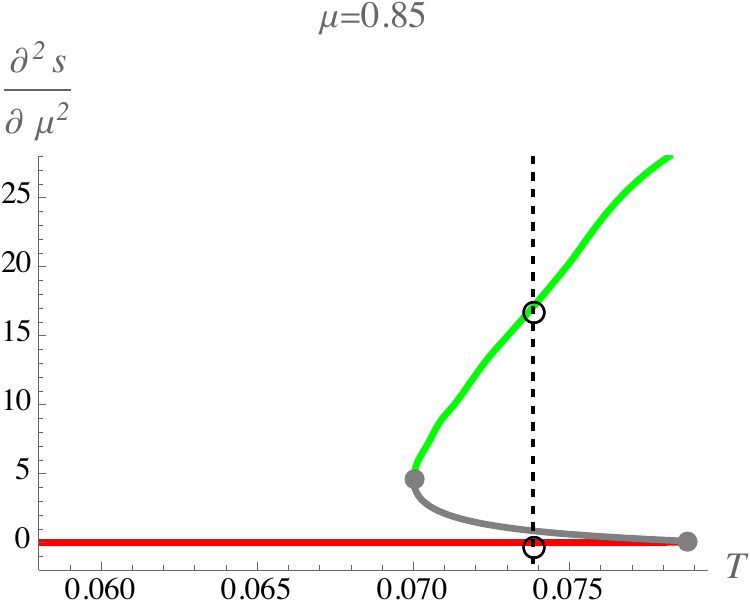}
    \vfill\vspace{10pt} 
    \includegraphics[width=0.325\linewidth]{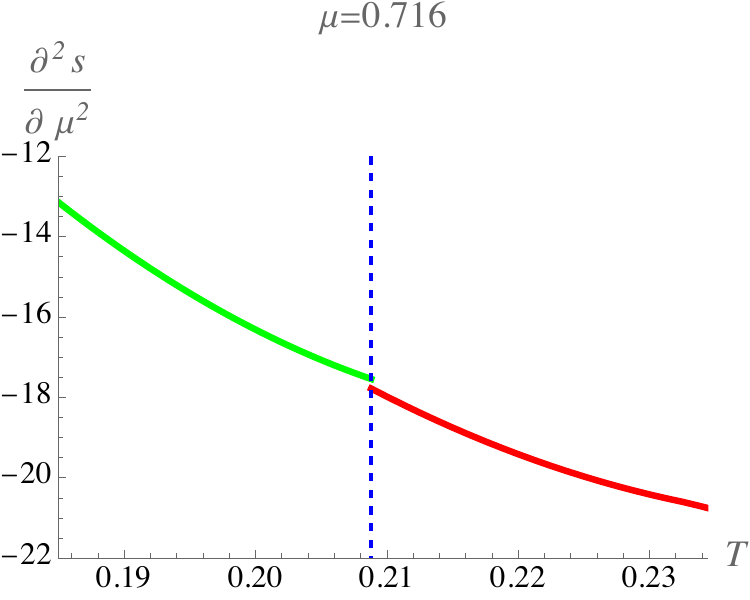}
    \includegraphics[width=0.325\linewidth]{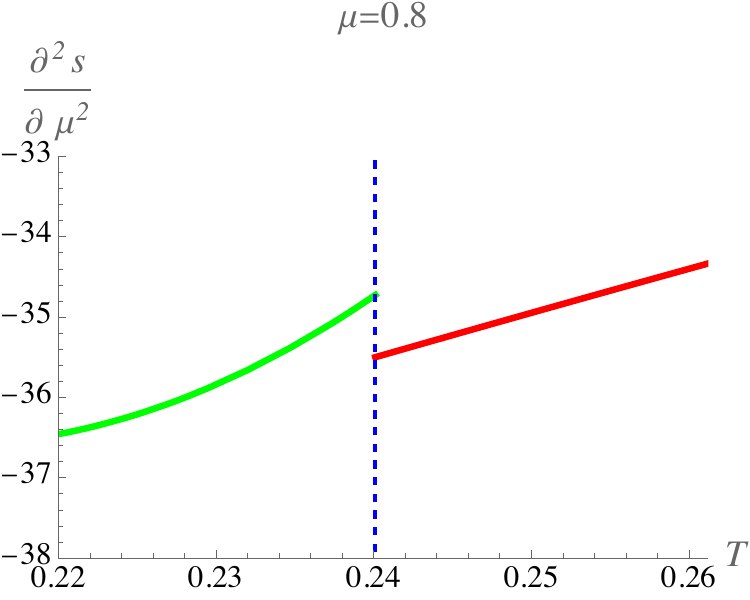}
    \includegraphics[width=0.325\linewidth]{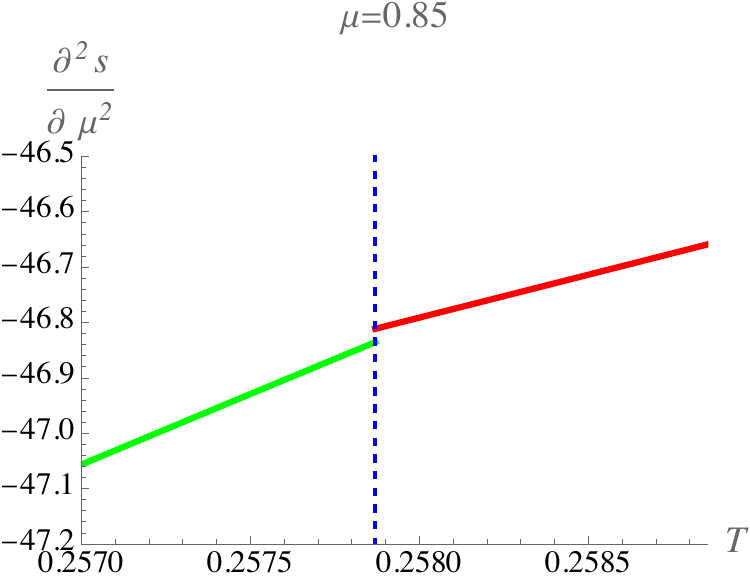}
    \caption{
    The same as Fig.~\ref{fig=freeenergy} but for the second-order derivative of the entropy density as a function of temperature for different chemical potentials.
    In the first row, the top-middle and top-right plots correspond to first-order transitions, where the derivative of the entropy density features a continuous ``S'' shape, leading to a finite vertical jump at the transition temperature between the two states indicated by empty black circles. 
    For the top-left plot and those in the second row, the evolutions of the entropy density are discontinuous, indicating that the transition is of the third order.
    }
    \label{fig=ddentropy-T}
\end{figure}

The resulting phase structure in terms of chemical potential and temperature is represented in Fig.~\ref{fig=phasediag}.
The region occupied by red dots represents the ultraviolet hairy black hole phase, which dominates most of the phase space.
The green triangles indicate the phase associated with the infrared hairy phase.
The transition between these two phases can be established using the Gibbs condition, as demonstrated in Figs.~\ref{fig=freeenergy} to~\ref{fig=ddentropy-T}.
As elaborated below, one encounters a first-order phase transition that terminates at a critical point, beyond which the curve turns back, and the transition becomes a third-order one. 
In Fig.~\ref{fig=phasediag}, the solid black curve presents the first-order phase transition, which is separated from a much smoother third-order phase transition denoted by the blue dashed curve by a critical point marked by a filled black circle.
It is noted that the bald black hole phase is not explicitly indicated on the phase diagram. 
This is because, according to Eq.~\eqref{eq=thermoquan}, the RN-AdS black hole discussed in Sec.~\ref{sec=model} corresponds to divergent temperature and chemical potential, effectively residing outside of the region of interest shown in Fig.~\ref{fig=phasediag}.

Based on the Ehrenfest classification, one evaluates the free energy density and its derivatives and explores the condition by equating the first-order derivatives, namely, the temperature and chemical potential, between the two phases.
The resulting free energy density and its derivatives are presented as functions of the chemical potential and temperature.
A first-order phase transition is observed, as shown in the first rows of Figs.~\ref{fig=freeenergy} and~\ref{fig=entropy-T}.
In Fig.~\ref{fig=freeenergy}, we present the free energy density as a function of temperature for different chemical potentials.
The first row shows three thermodynamic processes where a first-order transition occurs.
This is indicated by the zigzag trajectories in the phase space.
Specifically, the solid red curve represents states associated with the ultraviolet hairy black holes, while the solid green curve indicates those related to the infrared hairy black holes.
The solid blue circle indicates the point where the phase transition occurs. 
Beyond this point, the solid green and solid red curves correspond to the subcooled and superheated states that are metastable.
The solid gray curve represents hairy black hole solutions that are thermodynamically unstable.
The finiteness of the triangle closed by the red, green, and gray curves indicates that the transition is of first order.
As chemical potential decreases, the triangle continuously shrinks into a point indicated by the top-left plot.
This corresponds to the case where the system goes through the critical point.

These processes can also be analyzed in terms of the entropy density, 
the first derivative of the free energy density: $dF = -dP= - sdT +\mu d\rho$~\cite{Grefa:2021qvt}.
As shown in the first row of Fig.~\ref{fig=entropy-T}, for a first-order phase transition, the entropy density features a continuous ``S'' shape, which leads to a finite vertical jump at the transition temperature between the two states indicated by empty black circles.
Instead, it features a finite jump at the transition point while maintaining the chemical potential and temperature, reflecting the difference in the slopes between the solid red and solid green curves shown in the first row of Fig.~\ref{fig=freeenergy}.

Intriguingly, beyond the critical point, the transition curve does not disappear.
Specifically, the transition becomes the third order while the transition curve turns back as the chemical potential decreases, avoiding an intersection with the temperature axis.
The third-order transition curve is indicated by the dashed blue curve in Fig.~\ref{fig=phasediag}.
In our model, the above critical point is numerically found to be $(\mu_c, T_c)=(0.716,0.087)$, while the turning point is located at $(\mu_t, T_t)=(0.565, 0.125)$.
To demonstrate that there is indeed a third-order phase transition, one proceeds to evaluate the first and second-order derivatives of the entropy density, namely, the second and third-order derivatives of the free energy density.
The results are presented in Figs.~\ref{fig=dentropy-T} and~\ref{fig=ddentropy-T}.
Subsequently, the order of the phase transition beyond the critical point is determined according to the Ehrenfest classification, as demonstrated in Figs.~\ref{fig=freeenergy},~\ref{fig=entropy-T},~\ref{fig=dentropy-T}, and~\ref{fig=ddentropy-T}.
As shown in the second rows of Figs.~\ref{fig=freeenergy},~\ref{fig=entropy-T}, and~\ref{fig=dentropy-T}, the transition remains continuous until the second-order derivative of the Gibbs free energy density.
However, a finite jump is observed in the second row of Fig.~\ref{fig=ddentropy-T}, indicating that it is a third-order phase transition.
Numerically, such a transition is very smooth and elusive to observe.
However, the numerical results presented have reached an accuracy of one part in a thousand, which suffices to identify the observed discontinuity.
Physically, it might imply a minor break of the underlying symmetry, in accordance with the discussions about the field profiles regarding Fig.~\ref{fig=scahairBH2} below.
For comparison, concerning the first-order transition curve, a finite jump is always observed in all the derivatives of the free energy density, as shown in the top-middle and top-right plots in the first rows of Figs.~\ref{fig=entropy-T} to~\ref{fig=ddentropy-T}, confirming our findings.

Specifically, consider the evolution of a system with a given chemical potential more significant than that of the critical point $\mu>\mu_c$.
As the temperature drops continuously from a significant value, it will first suffer a third-order phase transition and then a first-order one.
However, a system with a smaller chemical potential but larger than the turning point $\mu_t<\mu<\mu_c$ will successively experience two third-order phase transitions.
Lastly, for a system with an even smaller chemical potential $\mu<\mu_t$, its evolution will be entirely continuous, corresponding to the {\it cross-over} in the low baryon density region.
Therefore, owing to the smooth transition at small chemical potentials and the elusive third-order phase transition, the present model is observationally consistent with the more prominent picture of the QCD phase diagram.



Before closing this section, we elaborate further on the profiles of the Maxwell field $\phi(r)$ and the metric functions $h(r)$, $A(r)$ with respect to the phase transition process.
In particular, we are interested in scrutinizing the difference in the profiles of the two hairy black holes as the phase transition occurs.
These results are presented in Fig.~\ref{fig=scahairBH2}.
We explore three scenarios corresponding to trajectories in the phase space with a given chemical potential that leads to a first-order phase transition, a third-order phase transition, and a process that intersects the critical point.
In the case of the first-order phase transition, the profiles are rather different on the two ends of the transition process.
This can be observed by comparing the red dashed curve with the green dashed one on the first row of Fig.~\ref{fig=scahairBH2}, representing an ultraviolet hairy black hole shortly before the transition and an infrared hairy one immediately after the transition.
On the other hand, as demonstrated in the middle and bottom rows, the region occupied by the red curves seamlessly merges with that filled with the green curves in a continuous fashion. 
Notably, the gap between the two phases becomes less apparent in the bottom row, representing the critical point, and the overall transition tends to exhibit a continuous behavior.
This indicates that the profiles are essentially indistinguishable in the case of the third-order phase transition and the trajectory intersecting the critical point.

\begin{figure}[thbp]
    \centering
    \includegraphics[width=0.3\linewidth]{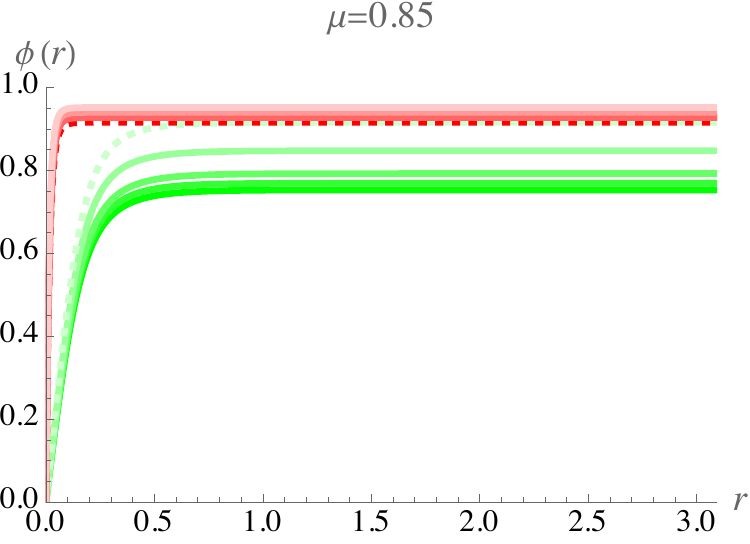}
    \includegraphics[width=0.3\linewidth]{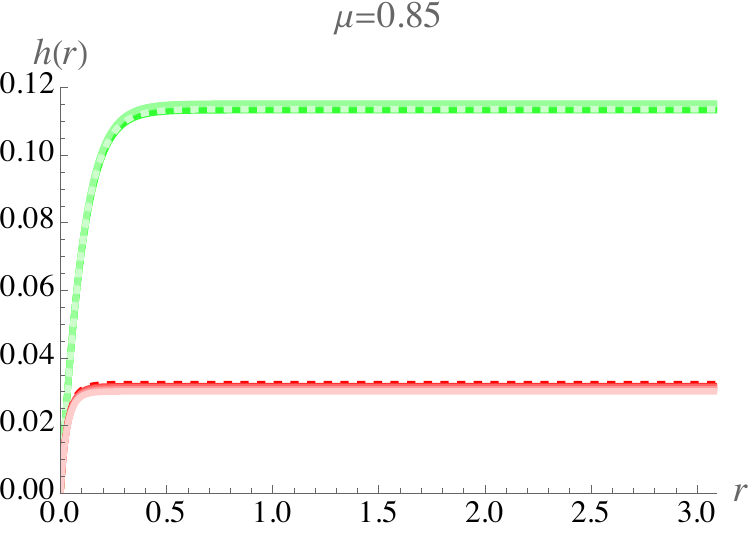}
    \includegraphics[width=0.38\linewidth]{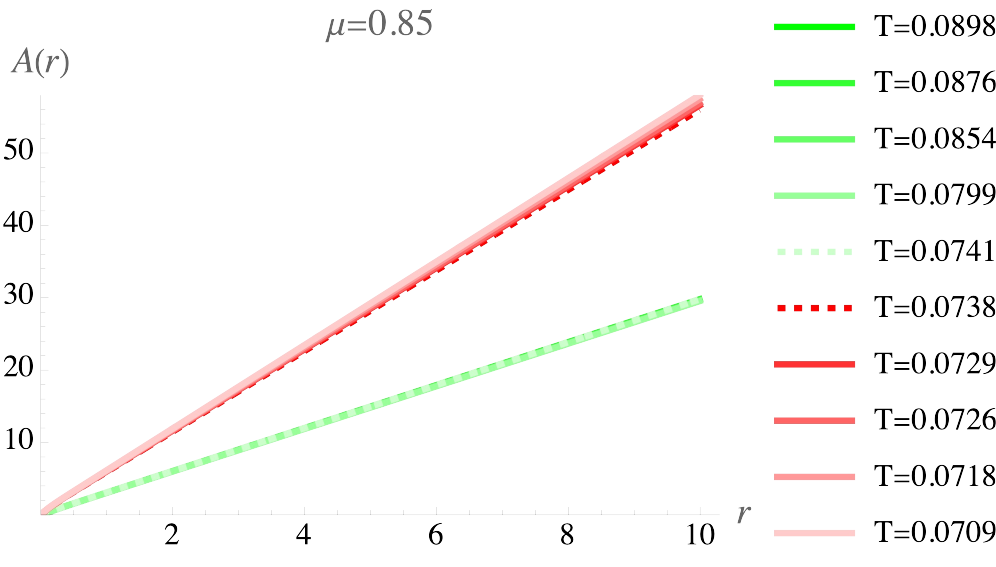}
    \vfill\vspace{10pt}
    \includegraphics[width=0.3\linewidth]{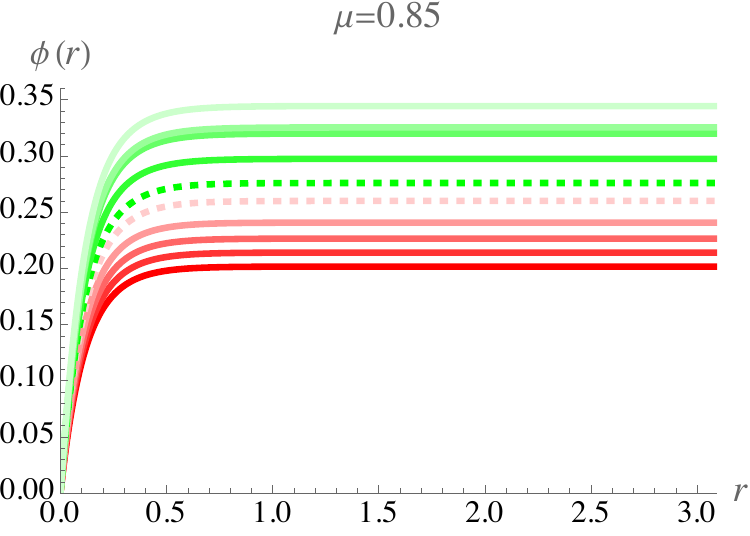}
    \includegraphics[width=0.3\linewidth]{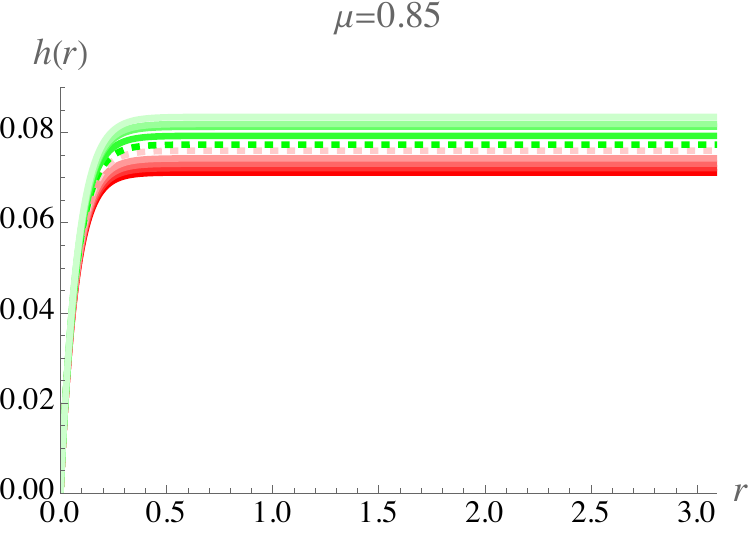}
    \includegraphics[width=0.38\linewidth]{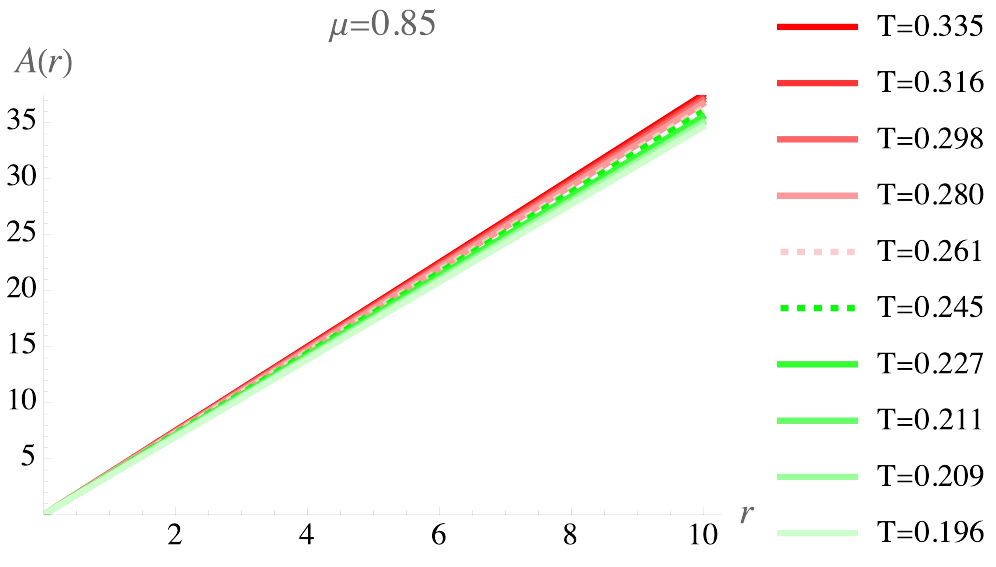}
    \vfill\vspace{10pt}
    \includegraphics[width=0.3\linewidth]{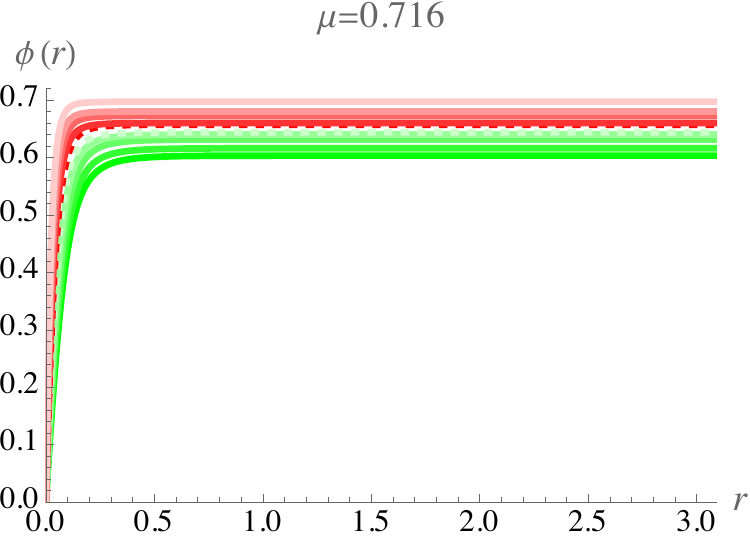}
    \includegraphics[width=0.3\linewidth]{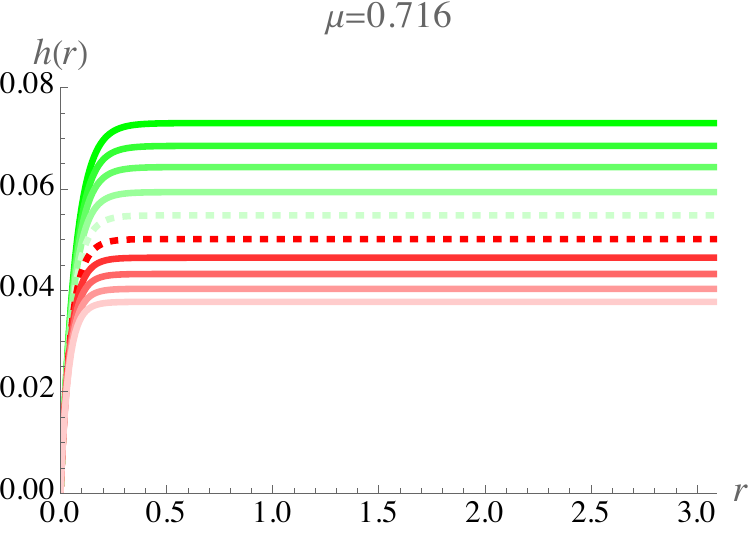}
    \includegraphics[width=0.38\linewidth]{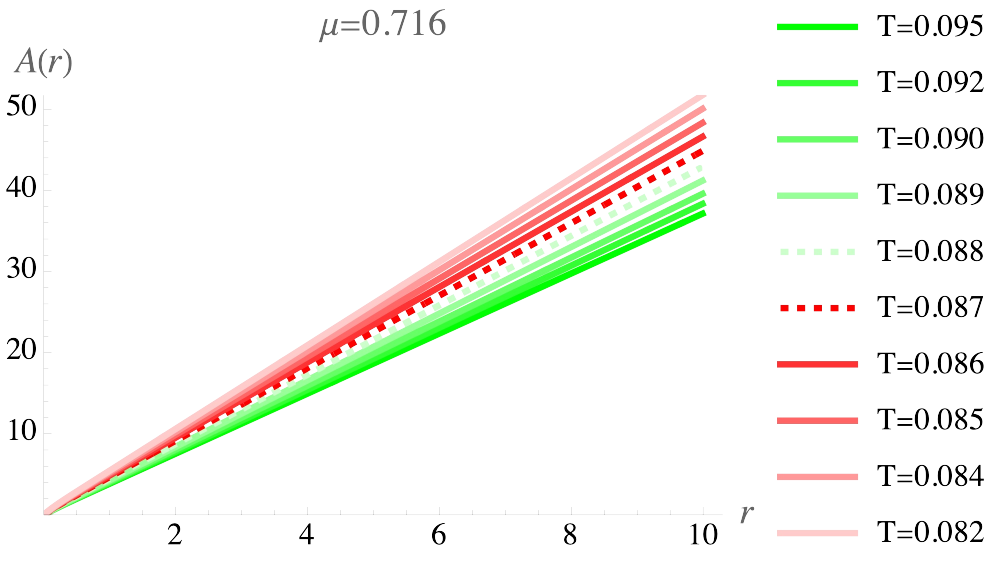}
    \caption{
    The profiles of the Maxwell field field $\phi(r)$ (left) and metric functions $h(r)$ (middle) $A(r)$ (right) associated with trajectories in the phase space.
    The red curves (with different line types) correspond to profiles of the ultraviolet hairy black holes evaluated at different temperatures but for a given chemical potential.
    Conversely, the green curves represent those of the infrared hairy black holes.
    Top row: Profiles associated with a process with $\mu=0.85$ that gives rise to a first-order phase transition between the two hairy black holes.
    Middle row: Profiles evaluated for a process with $\mu=0.85$ that leads to a third-order phase transition.
    Bottom row: Profiles calculated for a process with $\mu=0.716$ that intersects the critical point.}
    \label{fig=scahairBH2}
\end{figure}

\section{Concluding remarks}\label{sec=conclusion}

The holographic principle provides a means to utilize the black hole phase transitions in the bulk to study thermodynamic processes in the field theories on the AdS boundary. 
Such an approach has played a significant role in advancing theoretical physics over the past few decades. 
A prominent example is the holographic superconductor, which is triggered by the tachyonic instability below the critical temperature~\cite{Hartnoll:2008vx, Hartnoll:2008kx, Gubser:2008px}. 
The resulting second-order phase transition from a ``bald'' charged AdS black hole leads to the formation of a hairy black hole with scalar condensation in the bulk.
Intuitively, to mimic the QCD phase diagram, one needs the Maxwell degree freedom to furnish a finite chemical potential and a nonvanishing scalar hair to break the scaling invariance.
In other words, both phases must be scalarized hairy black holes.
Moreover, the two phases cannot be intrinsically different as they are expected to be indistinguishable, as the first-order phase transition is expected to terminate at a certain point in the phase diagram.
Therefore, exploring the properties and underlying instabilities in a theoretical framework with relevant degrees of freedom carried by the EMD~\cite{Cai:2022omk, Grefa:2021qvt, Critelli:2017oub} or, equivalently, the EMS~\cite{Herdeiro:2018wub, Fernandes:2019rez} theories is one of the primary motivations of the present study.

This work explores the black hole phase transitions and the subsequent phase diagram of an alternative EMD model.
On the one hand, while providing some insight into the QCD phase structure through the dilaton and Maxwell degrees of freedom, the construction of the model facilitates a transparent interpretation of the instabilities of the underlying black hole spacetimes and their transitions.
Compared to its counterpart implemented by the ESGB theory, the resulting phase structure exhibits unique properties that a first-order phase transition curve separating two phases whose difference eventually becomes hardly distinguishable, which differs significantly from previous results~\cite{Guo:2024vhq}.
On the other hand, when compared to the standard picture of the QCD phase diagram, the model indicates a somewhat different scenario.
Specifically, the first-order phase transition terminates at a critical point, beyond which the curve turns back, and the transition becomes third-order. 
Such a higher-order phase transition is numerically elusive, but our calculations have achieved satisfactory precision, ensuring its discrimination.
To ascertain our results, we explored the effective potential and profiles of the metric function, scalar, and Maxwell fields.
In particular, the properties of the phase diagram are closely scrutinized by numerically evaluating the Gibbs free energy and its derivatives using the AdS/CFT dictionary.
It is argued that while one encounters a critical point joining two types of phase transitions in the place of a critical endpoint, the model still offers a consistent interpretation for the observed cross-over in the low baryon density region.

The ongoing Beam Energy Scan (BES) program~\cite{RHIC-star-bes-01, RHIC-star-bes-03, RHIC-star-bes-05} at the RHIC (STAR) is dedicated to explore the QCD phase diagram.
Specifically, through two stages, BES-I and BES-II, it aims for Au+Au collisions from 3.0 to 7.7 GeV in the fixed-target mode and from 7.7 to 62.4 GeV in the collider mode.
Precise measurements have been realized for the high baryon density region of the QCD matter. 
The BES programs aim to map the size, shape, and temperature of the fireball produced in the collisions, as well as find the onset of deconfinement, signatures of a first-order phase transition, and a QCD critical point.
Intuitively, one might look for quantities sensitive to the underlying physics while accessible experimentally.
The higher cumulants of conserved charges and combinations of them, such as cumulant ratios, are candidates for such observables~\cite{statistical-model-03, statistical-model-04}.
These quantities fulfill the requirement, as they carry vital information on the primordial medium created in the collisions.
Moreover, it has been suggested~\cite{qcd-phase-fluctuations-review-02} that they are sensitive to the phase structure of the QCD matter and, in particular, the whereabouts of the critical point.
For the most part, the obtained results~\cite{statistical-model-10, statistical-model-08, urqmd-10, hydro-fluctuations-02,sph-bes-01} are qualitatively consistent with the experimental data~\cite{RHIC-star-bes-11,RHIC-star-bes-12,RHIC-star-bes-20, RHIC-star-bes-21}.
Although existing results have been inspiring~\cite{RHIC-star-bes-20, RHIC-star-bes-21, RHIC-star-bes-22, RHIC-star-bes-25}, to date, no decisive conclusion has been drawn about the whereabouts of the CEP, which leaves some room for further speculations.
In this regard, the present findings indicate a more complex QCD phase diagram that potentially implicates the search for a simple QCD critical point.

For the present study, we have chosen somewhat simplified forms for the scalar potential and the nonlinear interaction, and the model parameters have not been calibrated by matching to the lattice QCD data.
Nonetheless, for the most part, our results show remarkable consistency with the holographic QCD phase diagram obtained from the improved EMD model~\cite{Critelli:2017oub, Grefa:2021qvt}. 
Through instability analysis, we conclude that the emergence of such a phase structure can be understood from the perspective of hairy black holes in the bulk.
The phase transition is primarily governed by the interplay between the two tachyonic instabilities in the ultraviolet and infrared regions. 
It is understood that fine-tuning can be readily achieved by employing more complex functional forms of the potential and coupling functions in the action.
Such higher-order corrections substantially modify the tachyonic instabilities, but our findings indicate that the main features of the phase diagram are primarily determined at the leading order. 
In this regard, we speculate the possibility that the observed CEP in the standard EMD models might also extend further to a more elusive higher-order transition.
We plan to explore these aspects further in future studies.


\begin{acknowledgments}
We gratefully acknowledge the financial support from Brazilian agencies 
Funda\c{c}\~ao de Amparo \`a Pesquisa do Estado de S\~ao Paulo (FAPESP), 
Fundação de Amparo à Pesquisa do Estado do Rio Grande do Sul (FAPERGS),
Funda\c{c}\~ao de Amparo \`a Pesquisa do Estado do Rio de Janeiro (FAPERJ), 
Conselho Nacional de Desenvolvimento Cient\'{\i}fico e Tecnol\'ogico (CNPq), 
and Coordena\c{c}\~ao de Aperfei\c{c}oamento de Pessoal de N\'ivel Superior (CAPES).

\end{acknowledgments}

\begin{appendices}
\section{The instability analysis on effective potentials Eq.(23)}\label{sec=poten}

In this appendix, we investigate the properties of the effective potential in Eq.~\eqref{eq=potential} and analyze the tachyonic instabilities of the scalar perturbations.  
To facilitate the calculations, we utilize the scaling symmetry of the perturbed equation and set $L=1$ and $r_h=1$.
We will show that there are two types of tachyonic instabilities, which are controlled by the scalar mass $\lambda_V$ and the strength of the nonminimal coupling $\lambda_f$, respectively.

\begin{figure}[thbp]
    \centering
    \includegraphics[width=0.48\linewidth]{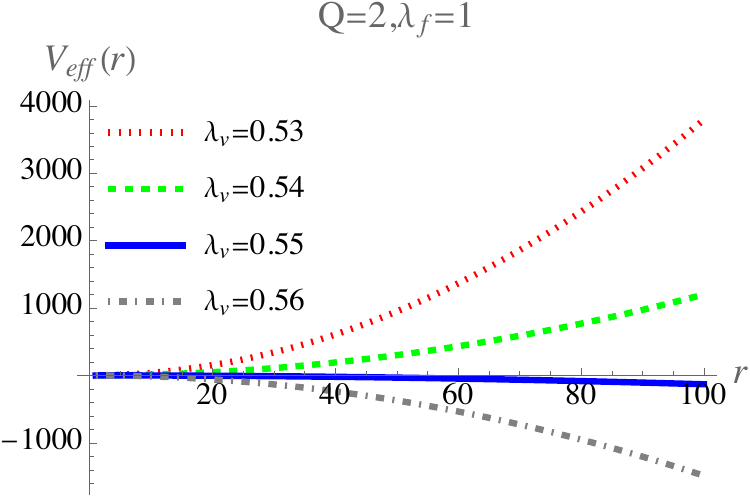}
    \includegraphics[width=0.48\linewidth]{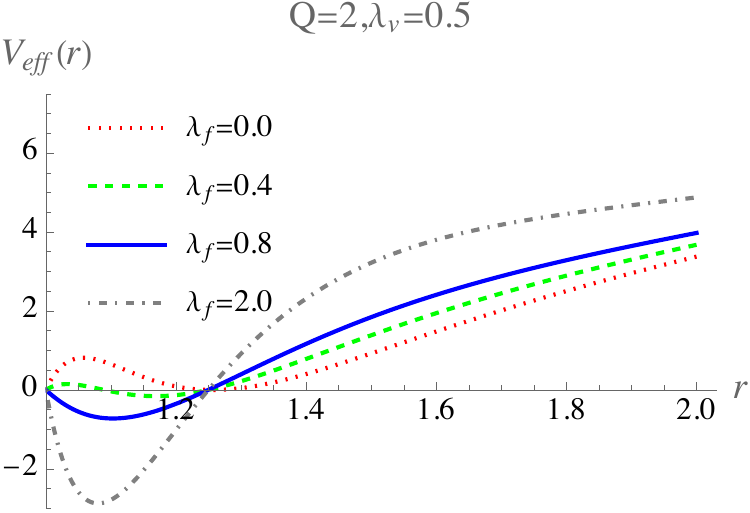}
    \caption{The profiles of the effective potential for different model parameters.
    Left: the effective potential for the tachyonic instability in the ultraviolet region with $Q=2, \lambda_f=1$, evaluated for different mass parameters $\lambda_V$.
    Right: the effective potential for the tachyonic instability in the infrared region with $Q=2, \lambda_V=0.5$, evaluated for different coupling parameters $\lambda_f$.}
    \label{fig=poten_lambda}
\end{figure}

The left panel of Fig.~\ref{fig=poten_lambda} shows how the mass parameter $\lambda_V$ triggers an instability at spatial infinity.
Specifically, for $Q=2$ and $\lambda_f=1$, the effective potential becomes negative in the ultraviolet region when $\lambda_V\gtrsim 0.55$.
Intuitively, such a negative region might host a bound state, destabilizing the ``bald'' charged black hole spacetime by continuously accumulating energy reminiscent of superradiant instability~\cite{Cardoso:2004nk, Brito:2015oca}.
Such a tachyonic instability gives rise to hairy black hole solutions, as first pointed out in~\cite{Hertog:2004bb}.
In the right panel, we demonstrate how the nonminimal coupling between the scalar field and the Maxwell field might induce a negative potential well near the event horizon.
For $Q=2$ and $\lambda_V=0.5$, the instability occurs when the coupling becomes strong enough $\lambda_f\gtrsim 0.4$.
This feature aligns with the spontaneous scalarization discussed by the ESGB theory~\cite{Myung:2018iyq, Guo:2020sdu}, where the tachyonic instability in the infrared region triggers a spontaneous scalarization.

Let us reiterate the above discussions in terms of the scalar field's effective mass, which possesses the form
\begin{equation}\label{eq=mass}
    m_{eff}^2=\frac{1}{4}F^2_{\mu\nu}f^0_{\psi}+V^0_{\psi}.
\end{equation}
In this expression, the second term on the r.h.s. of the equality comes from the second-order derivative of the potential function of the scalar field and is controlled by $\lambda_V$. 
Since $V^0_\psi < 0$ , it contributes negatively to the effective mass~\cite{Hartnoll:2008kx} and is referred to as ultraviolet instability in this work.
Conversely, near the horizon, the scalar field's effective potential becomes negative for large $\lambda_f$.
Such instability has been elaborated extensively in the literature in the context of spontaneous scalarization~\cite{Silva:2017uqg, Myung:2018iyq} and has been recognized mainly as tachyonic~\cite{buell1995potentials} (c.f.~\cite{Cardoso:2013opa, Cardoso:2013fwa}).
This second type of instability is referred to as infrared in the present study.
Regarding the effective mass Eq.~\eqref{eq=mass}, the parameter $\lambda_f$ governs the first term on the r.h.s. of the equality arising from the nonminimal coupling between the scalar field and the gauge field. 
Nonetheless, since $F^2_{\mu\nu} < 0$ and $f^0_\psi < 0$ as $\lambda_f>0$ in our paper, even though potentially overwhelmed by the second term, we note that this term may not always contribute negatively to the effective mass.

\begin{figure}[thbp]
    \centering
    \includegraphics[width=0.48\linewidth]{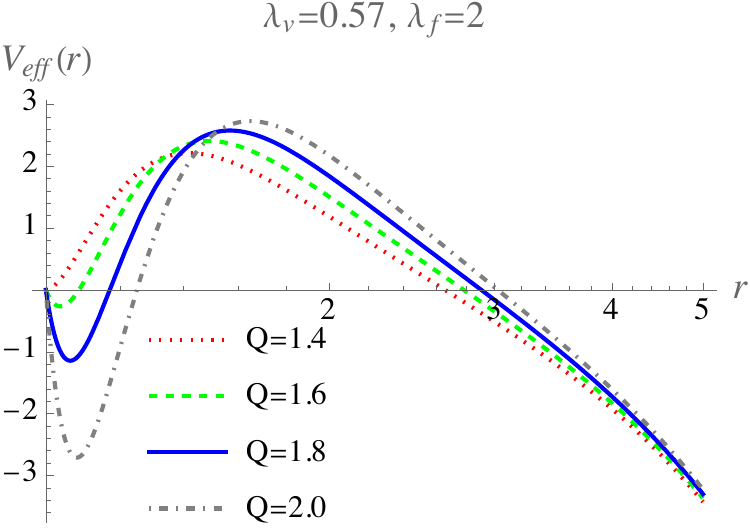}
    \caption{The profiles of the effective potential as a function of the black hole's charge $Q$.
    The calculations are carried out by assuming $\lambda_V=0.57$ and $\lambda_f=2$.}
    \label{fig=poten_Q}
\end{figure}

The above two types of tachyonic instabilities embedded in the present model could trigger two different hairy black holes.
So the transition among them potentially gives rise to a phase transition between the hairy black holes, as discussed in~\cite{Guo:2024vhq}.
To roughly understand the (phase) transition scenario, we point out an intriguing characteristic of the present model: the two types of instability discussed above can be controlled by a unique metric parameter, namely, the electric charge of the black hole $Q$.
Specifically, in Fig.~\ref{fig=poten_Q}, we illustrate that both tachyonic instabilities can be triggered by varying the black hole's charge while keeping the remaining model parameters fixed.
As shown in Fig.~\ref{fig=poten_Q}, a negative region in the effective potential is formed at large radial coordinates.
Moreover, as the black hole's charge increases, a potential well emerges near the horizon and becomes more significant, as indicated by the green dotted, blue solid, and gray dash-dotted curves.
This implies a competition between the two instabilities, typically leading to a first-order phase transition between the two underlying hairy black holes.
Furthermore, we can heuristically suggest that the two resulting hairy black holes might even become indistinguishable, as the parameters in the phase space are appropriately tuned, allowing the two regions of negative potential to merge seamlessly and lose their distinct identities.

\end{appendices}

\bibliographystyle{jhep}
\bibliography{references_guo, references_qian}

\providecommand{\href}[2]{#2}\begingroup\raggedright\begin{thebibliography}{10}

\bibitem{Harrison:2003sb}
M.~Harrison, T.~Ludlam and S.~Ozaki, \emph{{RHIC project overview}}, \href{http://dx.doi.org/10.1016/S0168-9002(02)01937-X}{\emph{Nucl. Instrum. Meth. A} {\bf 499} (2003) 235--244}.

\bibitem{2014APS..DNP.FJ001C}
D.~{Cebra} and {STAR Collaboration}, \emph{{Studying the Phase Diagram on QCD Matter at RHIC}},  in \emph{APS Division of Nuclear Physics Meeting Abstracts}, vol.~2014 of \emph{APS Meeting Abstracts}, p.~FJ.001, Sept., 2014.

\bibitem{Morrissey:2009tf}
D.~E. Morrissey, T.~Plehn and T.~M.~P. Tait, \emph{{Physics searches at the LHC}}, \href{http://dx.doi.org/10.1016/j.physrep.2012.02.007}{\emph{Phys. Rept.} {\bf 515} (2012) 1--113}, [\href{http://arxiv.org/abs/0912.3259}{{\tt 0912.3259}}].

\bibitem{Wells:2008xg}
J.~D. Wells, \emph{{How to Find a Hidden World at the Large Hadron Collider}},  \href{http://arxiv.org/abs/0803.1243}{{\tt 0803.1243}}.

\bibitem{Borsanyi:2010cj}
S.~Borsanyi, G.~Endrodi, Z.~Fodor, A.~Jakovac, S.~D. Katz, S.~Krieg et~al., \emph{{The QCD equation of state with dynamical quarks}}, \href{http://dx.doi.org/10.1007/JHEP11(2010)077}{\emph{JHEP} {\bf 11} (2010) 077}, [\href{http://arxiv.org/abs/1007.2580}{{\tt 1007.2580}}].

\bibitem{Borsanyi:2013bia}
S.~Borsanyi, Z.~Fodor, C.~Hoelbling, S.~D. Katz, S.~Krieg and K.~K. Szabo, \emph{{Full result for the QCD equation of state with 2+1 flavors}}, \href{http://dx.doi.org/10.1016/j.physletb.2014.01.007}{\emph{Phys. Lett. B} {\bf 730} (2014) 99--104}, [\href{http://arxiv.org/abs/1309.5258}{{\tt 1309.5258}}].

\bibitem{HotQCD:2014kol}
{\scshape HotQCD} collaboration, A.~Bazavov et~al., \emph{{Equation of state in ( 2+1 )-flavor QCD}}, \href{http://dx.doi.org/10.1103/PhysRevD.90.094503}{\emph{Phys. Rev. D} {\bf 90} (2014) 094503}, [\href{http://arxiv.org/abs/1407.6387}{{\tt 1407.6387}}].

\bibitem{Asakawa:1989bq}
M.~Asakawa and K.~Yazaki, \emph{{Chiral Restoration at Finite Density and Temperature}}, \href{http://dx.doi.org/10.1016/0375-9474(89)90002-X}{\emph{Nucl. Phys. A} {\bf 504} (1989) 668--684}.

\bibitem{Schwarz:1999dj}
T.~M. Schwarz, S.~P. Klevansky and G.~Papp, \emph{{The Phase diagram and bulk thermodynamical quantities in the NJL model at finite temperature and density}}, \href{http://dx.doi.org/10.1103/PhysRevC.60.055205}{\emph{Phys. Rev. C} {\bf 60} (1999) 055205}, [\href{http://arxiv.org/abs/nucl-th/9903048}{{\tt nucl-th/9903048}}].

\bibitem{Qin:2010nq}
S.-x. Qin, L.~Chang, H.~Chen, Y.-x. Liu and C.~D. Roberts, \emph{{Phase diagram and critical endpoint for strongly-interacting quarks}}, \href{http://dx.doi.org/10.1103/PhysRevLett.106.172301}{\emph{Phys. Rev. Lett.} {\bf 106} (2011) 172301}, [\href{http://arxiv.org/abs/1011.2876}{{\tt 1011.2876}}].

\bibitem{Shi:2014zpa}
C.~Shi, Y.-L. Wang, Y.~Jiang, Z.-F. Cui and H.-S. Zong, \emph{{Locate QCD Critical End Point in a Continuum Model Study}}, \href{http://dx.doi.org/10.1007/JHEP07(2014)014}{\emph{JHEP} {\bf 07} (2014) 014}, [\href{http://arxiv.org/abs/1403.3797}{{\tt 1403.3797}}].

\bibitem{Fu:2019hdw}
W.-j. Fu, J.~M. Pawlowski and F.~Rennecke, \emph{{QCD phase structure at finite temperature and density}}, \href{http://dx.doi.org/10.1103/PhysRevD.101.054032}{\emph{Phys. Rev. D} {\bf 101} (2020) 054032}, [\href{http://arxiv.org/abs/1909.02991}{{\tt 1909.02991}}].

\bibitem{Fu:2021oaw}
W.-j. Fu, X.~Luo, J.~M. Pawlowski, F.~Rennecke, R.~Wen and S.~Yin, \emph{{Hyper-order baryon number fluctuations at finite temperature and density}}, \href{http://dx.doi.org/10.1103/PhysRevD.104.094047}{\emph{Phys. Rev. D} {\bf 104} (2021) 094047}, [\href{http://arxiv.org/abs/2101.06035}{{\tt 2101.06035}}].

\bibitem{adscft-qcd-03}
J.~Erlich, E.~Katz, D.~T. Son and M.~A. Stephanov, \emph{{QCD and a holographic model of hadrons}}, \href{http://dx.doi.org/10.1103/PhysRevLett.95.261602}{\emph{Phys. Rev. Lett.} {\bf 95} (2005) 261602}, [\href{http://arxiv.org/abs/hep-ph/0501128}{{\tt hep-ph/0501128}}].

\bibitem{adscft-qcd-05}
N.~Horigome and Y.~Tanii, \emph{{Holographic chiral phase transition with chemical potential}}, \href{http://dx.doi.org/10.1088/1126-6708/2007/01/072}{\emph{JHEP} {\bf 01} (2007) 072}, [\href{http://arxiv.org/abs/hep-th/0608198}{{\tt hep-th/0608198}}].

\bibitem{adscft-qcd-07}
U.~Gursoy and E.~Kiritsis, \emph{{Exploring improved holographic theories for QCD: Part I}}, \href{http://dx.doi.org/10.1088/1126-6708/2008/02/032}{\emph{JHEP} {\bf 02} (2008) 032}, [\href{http://arxiv.org/abs/0707.1324}{{\tt 0707.1324}}].

\bibitem{adscft-qcd-08}
U.~Gursoy, E.~Kiritsis and F.~Nitti, \emph{{Exploring improved holographic theories for QCD: Part II}}, \href{http://dx.doi.org/10.1088/1126-6708/2008/02/019}{\emph{JHEP} {\bf 02} (2008) 019}, [\href{http://arxiv.org/abs/arX0707.1349}{{\tt arX0707.1349}}].

\bibitem{adscft-qcd-emd-10}
D.~Dudal and S.~Mahapatra, \emph{{Thermal entropy of a quark-antiquark pair above and below deconfinement from a dynamical holographic QCD model}}, \href{http://dx.doi.org/10.1103/PhysRevD.96.126010}{\emph{Phys. Rev. D} {\bf 96} (2017) 126010}, [\href{http://arxiv.org/abs/1708.06995}{{\tt 1708.06995}}].

\bibitem{Kim:2012ey}
Y.~Kim, I.~J. Shin and T.~Tsukioka, \emph{{Holographic QCD: Past, Present, and Future}}, \href{http://dx.doi.org/10.1016/j.ppnp.2012.09.002}{\emph{Prog. Part. Nucl. Phys.} {\bf 68} (2013) 55--112}, [\href{http://arxiv.org/abs/1205.4852}{{\tt 1205.4852}}].

\bibitem{Cai:2012xh}
R.-G. Cai, S.~He and D.~Li, \emph{{A hQCD model and its phase diagram in Einstein-Maxwell-Dilaton system}}, \href{http://dx.doi.org/10.1007/JHEP03(2012)033}{\emph{JHEP} {\bf 03} (2012) 033}, [\href{http://arxiv.org/abs/1201.0820}{{\tt 1201.0820}}].

\bibitem{DeWolfe:2010he}
O.~DeWolfe, S.~S. Gubser and C.~Rosen, \emph{{A holographic critical point}}, \href{http://dx.doi.org/10.1103/PhysRevD.83.086005}{\emph{Phys. Rev. D} {\bf 83} (2011) 086005}, [\href{http://arxiv.org/abs/1012.1864}{{\tt 1012.1864}}].

\bibitem{adscft-qcd-40}
X.~Chen, D.~Li, D.~Hou and M.~Huang, \emph{{Quarkyonic phase from quenched dynamical holographic QCD model}}, \href{http://dx.doi.org/10.1007/JHEP03(2020)073}{\emph{JHEP} {\bf 03} (2020) 073}, [\href{http://arxiv.org/abs/1908.02000}{{\tt 1908.02000}}].

\bibitem{Erdmenger:2020flu}
J.~Erdmenger, N.~Evans, W.~Porod and K.~S. Rigatos, \emph{{Gauge/gravity dual dynamics for the strongly coupled sector of composite Higgs models}}, \href{http://dx.doi.org/10.1007/JHEP02(2021)058}{\emph{JHEP} {\bf 02} (2021) 058}, [\href{http://arxiv.org/abs/2010.10279}{{\tt 2010.10279}}].

\bibitem{DeWolfe:2011ts}
O.~DeWolfe, S.~S. Gubser and C.~Rosen, \emph{{Dynamic critical phenomena at a holographic critical point}}, \href{http://dx.doi.org/10.1103/PhysRevD.84.126014}{\emph{Phys. Rev. D} {\bf 84} (2011) 126014}, [\href{http://arxiv.org/abs/1108.2029}{{\tt 1108.2029}}].

\bibitem{adscft-qcd-emd-11}
X.~Chen, D.~Li, D.~Hou and M.~Huang, \emph{{Quarkyonic phase from quenched dynamical holographic QCD model}}, \href{http://dx.doi.org/10.1007/JHEP03(2020)073}{\emph{JHEP} {\bf 03} (2020) 073}, [\href{http://arxiv.org/abs/1908.02000}{{\tt 1908.02000}}].

\bibitem{adscft-qcd-emd-12}
X.~Chen, L.~Zhang, D.~Li, D.~Hou and M.~Huang, \emph{{Gluodynamics and deconfinement phase transition under rotation from holography}}, \href{http://dx.doi.org/10.1007/JHEP07(2021)132}{\emph{JHEP} {\bf 07} (2021) 132}, [\href{http://arxiv.org/abs/2010.14478}{{\tt 2010.14478}}].

\bibitem{Chen:2022goa}
Y.~Chen, D.~Li and M.~Huang, \emph{{The dynamical holographic QCD method for hadron physics and QCD matter}}, \href{http://dx.doi.org/10.1088/1572-9494/ac82ad}{\emph{Commun. Theor. Phys.} {\bf 74} (2022) 097201}, [\href{http://arxiv.org/abs/2206.00917}{{\tt 2206.00917}}].

\bibitem{Fu:2024wkn}
Q.~Fu, S.~He, L.~Li and Z.~Li, \emph{{Revisiting holographic model for thermal and dense QCD with a critical point}},  \href{http://arxiv.org/abs/2404.12109}{{\tt 2404.12109}}.

\bibitem{Li:2024lrh}
Z.~Li, \emph{{Universality of holographic entanglement properties in holographic QCD}},  \href{http://arxiv.org/abs/2402.02944}{{\tt 2402.02944}}.

\bibitem{Critelli:2017oub}
R.~Critelli, J.~Noronha, J.~Noronha-Hostler, I.~Portillo, C.~Ratti and R.~Rougemont, \emph{{Critical point in the phase diagram of primordial quark-gluon matter from black hole physics}}, \href{http://dx.doi.org/10.1103/PhysRevD.96.096026}{\emph{Phys. Rev. D} {\bf 96} (2017) 096026}, [\href{http://arxiv.org/abs/1706.00455}{{\tt 1706.00455}}].

\bibitem{Grefa:2021qvt}
J.~Grefa, J.~Noronha, J.~Noronha-Hostler, I.~Portillo, C.~Ratti and R.~Rougemont, \emph{{Hot and dense quark-gluon plasma thermodynamics from holographic black holes}}, \href{http://dx.doi.org/10.1103/PhysRevD.104.034002}{\emph{Phys. Rev. D} {\bf 104} (2021) 034002}, [\href{http://arxiv.org/abs/2102.12042}{{\tt 2102.12042}}].

\bibitem{Knaute:2017opk}
J.~Knaute, R.~Yaresko and B.~K\"ampfer, \emph{{Holographic QCD phase diagram with critical point from Einstein\textendash{}Maxwell-dilaton dynamics}}, \href{http://dx.doi.org/10.1016/j.physletb.2018.01.053}{\emph{Phys. Lett. B} {\bf 778} (2018) 419--425}, [\href{http://arxiv.org/abs/1702.06731}{{\tt 1702.06731}}].

\bibitem{Cai:2022omk}
R.-G. Cai, S.~He, L.~Li and Y.-X. Wang, \emph{{Probing QCD critical point and induced gravitational wave by black hole physics}}, \href{http://dx.doi.org/10.1103/PhysRevD.106.L121902}{\emph{Phys. Rev. D} {\bf 106} (2022) L121902}, [\href{http://arxiv.org/abs/2201.02004}{{\tt 2201.02004}}].

\bibitem{Chen:2024ckb}
X.~Chen and M.~Huang, \emph{{Machine learning holographic black hole from lattice QCD equation of state}}, \href{http://dx.doi.org/10.1103/PhysRevD.109.L051902}{\emph{Phys. Rev. D} {\bf 109} (2024) L051902}, [\href{http://arxiv.org/abs/2401.06417}{{\tt 2401.06417}}].

\bibitem{Liu:2023pbt}
X.-Y. Liu, X.-C. Peng, Y.-L. Wu and Z.~Fang, \emph{{Holographic study on QCD phase transition and phase diagram with two flavors}}, \href{http://dx.doi.org/10.1103/PhysRevD.109.054032}{\emph{Phys. Rev. D} {\bf 109} (2024) 054032}, [\href{http://arxiv.org/abs/2312.01346}{{\tt 2312.01346}}].

\bibitem{Sudarsky:2002mk}
D.~Sudarsky and J.~A. Gonzalez, \emph{{On black hole scalar hair in asymptotically anti-de Sitter space-times}}, \href{http://dx.doi.org/10.1103/PhysRevD.67.024038}{\emph{Phys. Rev. D} {\bf 67} (2003) 024038}, [\href{http://arxiv.org/abs/gr-qc/0207069}{{\tt gr-qc/0207069}}].

\bibitem{Ruffini:1971bza}
R.~Ruffini and J.~A. Wheeler, \emph{{Introducing the black hole}}, \href{http://dx.doi.org/10.1063/1.3022513}{\emph{Phys. Today} {\bf 24} (1971) 30}.

\bibitem{Chrusciel:2012jk}
P.~T. Chrusciel, J.~Lopes~Costa and M.~Heusler, \emph{{Stationary Black Holes: Uniqueness and Beyond}}, \href{http://dx.doi.org/10.12942/lrr-2012-7}{\emph{Living Rev. Rel.} {\bf 15} (2012) 7}, [\href{http://arxiv.org/abs/1205.6112}{{\tt 1205.6112}}].

\bibitem{Carter:1971zc}
B.~Carter, \emph{{Axisymmetric Black Hole Has Only Two Degrees of Freedom}}, \href{http://dx.doi.org/10.1103/PhysRevLett.26.331}{\emph{Phys. Rev. Lett.} {\bf 26} (1971) 331--333}.

\bibitem{Gubser:2008px}
S.~S. Gubser, \emph{{Breaking an Abelian gauge symmetry near a black hole horizon}}, \href{http://dx.doi.org/10.1103/PhysRevD.78.065034}{\emph{Phys. Rev. D} {\bf 78} (2008) 065034}, [\href{http://arxiv.org/abs/0801.2977}{{\tt 0801.2977}}].

\bibitem{Hartnoll:2008kx}
S.~A. Hartnoll, C.~P. Herzog and G.~T. Horowitz, \emph{{Holographic Superconductors}}, \href{http://dx.doi.org/10.1088/1126-6708/2008/12/015}{\emph{JHEP} {\bf 12} (2008) 015}, [\href{http://arxiv.org/abs/0810.1563}{{\tt 0810.1563}}].

\bibitem{Antoniou:2017acq}
G.~Antoniou, A.~Bakopoulos and P.~Kanti, \emph{{Evasion of No-Hair Theorems and Novel Black-Hole Solutions in Gauss-Bonnet Theories}}, \href{http://dx.doi.org/10.1103/PhysRevLett.120.131102}{\emph{Phys. Rev. Lett.} {\bf 120} (2018) 131102}, [\href{http://arxiv.org/abs/1711.03390}{{\tt 1711.03390}}].

\bibitem{Doneva:2017bvd}
D.~D. Doneva and S.~S. Yazadjiev, \emph{{New Gauss-Bonnet Black Holes with Curvature-Induced Scalarization in Extended Scalar-Tensor Theories}}, \href{http://dx.doi.org/10.1103/PhysRevLett.120.131103}{\emph{Phys. Rev. Lett.} {\bf 120} (2018) 131103}, [\href{http://arxiv.org/abs/1711.01187}{{\tt 1711.01187}}].

\bibitem{Silva:2017uqg}
H.~O. Silva, J.~Sakstein, L.~Gualtieri, T.~P. Sotiriou and E.~Berti, \emph{{Spontaneous scalarization of black holes and compact stars from a Gauss-Bonnet coupling}}, \href{http://dx.doi.org/10.1103/PhysRevLett.120.131104}{\emph{Phys. Rev. Lett.} {\bf 120} (2018) 131104}, [\href{http://arxiv.org/abs/1711.02080}{{\tt 1711.02080}}].

\bibitem{Blazquez-Salcedo:2018jnn}
J.~L. Bl\'azquez-Salcedo, D.~D. Doneva, J.~Kunz and S.~S. Yazadjiev, \emph{{Radial perturbations of the scalarized Einstein-Gauss-Bonnet black holes}}, \href{http://dx.doi.org/10.1103/PhysRevD.98.084011}{\emph{Phys. Rev. D} {\bf 98} (2018) 084011}, [\href{http://arxiv.org/abs/1805.05755}{{\tt 1805.05755}}].

\bibitem{Myung:2018iyq}
Y.~S. Myung and D.-C. Zou, \emph{{Gregory-Laflamme instability of black hole in Einstein-scalar-Gauss-Bonnet theories}}, \href{http://dx.doi.org/10.1103/PhysRevD.98.024030}{\emph{Phys. Rev. D} {\bf 98} (2018) 024030}, [\href{http://arxiv.org/abs/1805.05023}{{\tt 1805.05023}}].

\bibitem{Cardoso:2013fwa}
V.~Cardoso, I.~P. Carucci, P.~Pani and T.~P. Sotiriou, \emph{{Black holes with surrounding matter in scalar-tensor theories}}, \href{http://dx.doi.org/10.1103/PhysRevLett.111.111101}{\emph{Phys. Rev. Lett.} {\bf 111} (2013) 111101}, [\href{http://arxiv.org/abs/1308.6587}{{\tt 1308.6587}}].

\bibitem{buell1995potentials}
W.~F. Buell and B.~Shadwick, \emph{Potentials and bound states}, {\emph{American Journal of Physics} {\bf 63} (1995) 256--258}.

\bibitem{Cardoso:2013opa}
V.~Cardoso, I.~P. Carucci, P.~Pani and T.~P. Sotiriou, \emph{{Matter around Kerr black holes in scalar-tensor theories: scalarization and superradiant instability}}, \href{http://dx.doi.org/10.1103/PhysRevD.88.044056}{\emph{Phys. Rev. D} {\bf 88} (2013) 044056}, [\href{http://arxiv.org/abs/1305.6936}{{\tt 1305.6936}}].

\bibitem{Brigante:2008gz}
M.~Brigante, H.~Liu, R.~C. Myers, S.~Shenker and S.~Yaida, \emph{{The Viscosity Bound and Causality Violation}}, \href{http://dx.doi.org/10.1103/PhysRevLett.100.191601}{\emph{Phys. Rev. Lett.} {\bf 100} (2008) 191601}, [\href{http://arxiv.org/abs/0802.3318}{{\tt 0802.3318}}].

\bibitem{Brigante:2007nu}
M.~Brigante, H.~Liu, R.~C. Myers, S.~Shenker and S.~Yaida, \emph{{Viscosity Bound Violation in Higher Derivative Gravity}}, \href{http://dx.doi.org/10.1103/PhysRevD.77.126006}{\emph{Phys. Rev. D} {\bf 77} (2008) 126006}, [\href{http://arxiv.org/abs/0712.0805}{{\tt 0712.0805}}].

\bibitem{Buchel:2009tt}
A.~Buchel and R.~C. Myers, \emph{{Causality of Holographic Hydrodynamics}}, \href{http://dx.doi.org/10.1088/1126-6708/2009/08/016}{\emph{JHEP} {\bf 08} (2009) 016}, [\href{http://arxiv.org/abs/0906.2922}{{\tt 0906.2922}}].

\bibitem{Hofman:2008ar}
D.~M. Hofman and J.~Maldacena, \emph{{Conformal collider physics: Energy and charge correlations}}, \href{http://dx.doi.org/10.1088/1126-6708/2008/05/012}{\emph{JHEP} {\bf 05} (2008) 012}, [\href{http://arxiv.org/abs/0803.1467}{{\tt 0803.1467}}].

\bibitem{Hofman:2009ug}
D.~M. Hofman, \emph{{Higher Derivative Gravity, Causality and Positivity of Energy in a UV complete QFT}}, \href{http://dx.doi.org/10.1016/j.nuclphysb.2009.08.001}{\emph{Nucl. Phys. B} {\bf 823} (2009) 174--194}, [\href{http://arxiv.org/abs/0907.1625}{{\tt 0907.1625}}].

\bibitem{Camanho:2009vw}
X.~O. Camanho and J.~D. Edelstein, \emph{{Causality constraints in AdS/CFT from conformal collider physics and Gauss-Bonnet gravity}}, \href{http://dx.doi.org/10.1007/JHEP04(2010)007}{\emph{JHEP} {\bf 04} (2010) 007}, [\href{http://arxiv.org/abs/0911.3160}{{\tt 0911.3160}}].

\bibitem{Herdeiro:2018wub}
C.~A.~R. Herdeiro, E.~Radu, N.~Sanchis-Gual and J.~A. Font, \emph{{Spontaneous Scalarization of Charged Black Holes}}, \href{http://dx.doi.org/10.1103/PhysRevLett.121.101102}{\emph{Phys. Rev. Lett.} {\bf 121} (2018) 101102}, [\href{http://arxiv.org/abs/1806.05190}{{\tt 1806.05190}}].

\bibitem{Fernandes:2019rez}
P.~G.~S. Fernandes, C.~A.~R. Herdeiro, A.~M. Pombo, E.~Radu and N.~Sanchis-Gual, \emph{{Spontaneous Scalarisation of Charged Black Holes: Coupling Dependence and Dynamical Features}}, \href{http://dx.doi.org/10.1088/1361-6382/ab23a1}{\emph{Class. Quant. Grav.} {\bf 36} (2019) 134002}, [\href{http://arxiv.org/abs/1902.05079}{{\tt 1902.05079}}].

\bibitem{Zou:2019bpt}
D.-C. Zou and Y.~S. Myung, \emph{{Scalarized charged black holes with scalar mass term}}, \href{http://dx.doi.org/10.1103/PhysRevD.100.124055}{\emph{Phys. Rev. D} {\bf 100} (2019) 124055}, [\href{http://arxiv.org/abs/1909.11859}{{\tt 1909.11859}}].

\bibitem{Guo:2021zed}
G.~Guo, P.~Wang, H.~Wu and H.~Yang, \emph{{Scalarized Einstein\textendash{}Maxwell-scalar black holes in anti-de Sitter spacetime}}, \href{http://dx.doi.org/10.1140/epjc/s10052-021-09614-7}{\emph{Eur. Phys. J. C} {\bf 81} (2021) 864}, [\href{http://arxiv.org/abs/2102.04015}{{\tt 2102.04015}}].

\bibitem{Deshpande:2024itz}
R.~Deshpande and O.~Lunin, \emph{{Multi-charged geometries with cosmological constant}},  \href{http://arxiv.org/abs/2408.08254}{{\tt 2408.08254}}.

\bibitem{Hertog:2004bb}
T.~Hertog and K.~Maeda, \emph{{Stability and thermodynamics of AdS black holes with scalar hair}}, \href{http://dx.doi.org/10.1103/PhysRevD.71.024001}{\emph{Phys. Rev. D} {\bf 71} (2005) 024001}, [\href{http://arxiv.org/abs/hep-th/0409314}{{\tt hep-th/0409314}}].

\bibitem{Perivolaropoulos:2020uqy}
L.~Perivolaropoulos and F.~Skara, \emph{{Scalar tachyonic instabilities in gravitational backgrounds: Existence and growth rate}}, \href{http://dx.doi.org/10.1103/PhysRevD.102.104034}{\emph{Phys. Rev. D} {\bf 102} (2020) 104034}, [\href{http://arxiv.org/abs/2009.05640}{{\tt 2009.05640}}].

\bibitem{Guo:2020sdu}
H.~Guo, S.~Kiorpelidi, X.-M. Kuang, E.~Papantonopoulos, B.~Wang and J.-P. Wu, \emph{{Spontaneous holographic scalarization of black holes in Einstein-scalar-Gauss-Bonnet theories}}, \href{http://dx.doi.org/10.1103/PhysRevD.102.084029}{\emph{Phys. Rev. D} {\bf 102} (2020) 084029}, [\href{http://arxiv.org/abs/2006.10659}{{\tt 2006.10659}}].

\bibitem{Guo:2024vhq}
H.~Guo, W.-L. Qian and B.~Wang, \emph{{Phase structure of holographic superconductors in an Einstein-scalar-Gauss-Bonnet theory with spontaneous scalarization}}, \href{http://dx.doi.org/10.1103/PhysRevD.109.124038}{\emph{Phys. Rev. D} {\bf 109} (2024) 124038}, [\href{http://arxiv.org/abs/2401.09846}{{\tt 2401.09846}}].

\bibitem{Kubiznak:2016qmn}
D.~Kubiznak, R.~B. Mann and M.~Teo, \emph{{Black hole chemistry: thermodynamics with Lambda}}, \href{http://dx.doi.org/10.1088/1361-6382/aa5c69}{\emph{Class. Quant. Grav.} {\bf 34} (2017) 063001}, [\href{http://arxiv.org/abs/1608.06147}{{\tt 1608.06147}}].

\bibitem{Gubser:2008zu}
S.~S. Gubser, \emph{{Colorful horizons with charge in anti-de Sitter space}}, \href{http://dx.doi.org/10.1103/PhysRevLett.101.191601}{\emph{Phys. Rev. Lett.} {\bf 101} (2008) 191601}, [\href{http://arxiv.org/abs/0803.3483}{{\tt 0803.3483}}].

\bibitem{Hertog:2004dr}
T.~Hertog and K.~Maeda, \emph{{Black holes with scalar hair and asymptotics in N = 8 supergravity}}, \href{http://dx.doi.org/10.1088/1126-6708/2004/07/051}{\emph{JHEP} {\bf 07} (2004) 051}, [\href{http://arxiv.org/abs/hep-th/0404261}{{\tt hep-th/0404261}}].

\bibitem{Torii:2001pg}
T.~Torii, K.~Maeda and M.~Narita, \emph{{Scalar hair on the black hole in asymptotically anti-de Sitter space-time}}, \href{http://dx.doi.org/10.1103/PhysRevD.64.044007}{\emph{Phys. Rev. D} {\bf 64} (2001) 044007}.

\bibitem{Hartnoll:2008vx}
S.~A. Hartnoll, C.~P. Herzog and G.~T. Horowitz, \emph{{Building a Holographic Superconductor}}, \href{http://dx.doi.org/10.1103/PhysRevLett.101.031601}{\emph{Phys. Rev. Lett.} {\bf 101} (2008) 031601}, [\href{http://arxiv.org/abs/0803.3295}{{\tt 0803.3295}}].

\bibitem{RHIC-star-bes-01}
{\scshape STAR} collaboration, B.~Mohanty, \emph{{STAR experiment results from the beam energy scan program at RHIC}}, \href{http://dx.doi.org/10.1088/0954-3899/38/12/124023}{\emph{J. Phys.} {\bf G38} (2011) 124023}, [\href{http://arxiv.org/abs/1106.5902}{{\tt 1106.5902}}].

\bibitem{RHIC-star-bes-03}
{\scshape STAR} collaboration, L.~Kumar, \emph{{STAR Results from the RHIC Beam Energy Scan-I}}, \href{http://dx.doi.org/10.1016/j.nuclphysa.2013.01.070}{\emph{Nucl. Phys.} {\bf A904-905} (2013) 256c--263c}, [\href{http://arxiv.org/abs/1211.1350}{{\tt 1211.1350}}].

\bibitem{RHIC-star-bes-05}
{\scshape STAR} collaboration, C.~Yang, \emph{{The STAR beam energy scan phase II physics and upgrades}}, \href{http://dx.doi.org/10.1016/j.nuclphysa.2017.05.042}{\emph{Nucl. Phys.} {\bf A967} (2017) 800--803}.

\bibitem{statistical-model-03}
F.~Becattini, A.~Keranen, L.~Ferroni and T.~Gabbriellini, \emph{{Multiplicity fluctuations in the hadron gas with exact conservation laws}}, \href{http://dx.doi.org/10.1103/PhysRevC.72.064904}{\emph{Phys. Rev.} {\bf C72} (2005) 064904}, [\href{http://arxiv.org/abs/nucl-th/0507039}{{\tt nucl-th/0507039}}].

\bibitem{statistical-model-04}
V.~Begun, M.~I. Gorenstein, M.~Hauer, V.~Konchakovski and O.~Zozulya, \emph{{Multiplicity Fluctuations in Hadron-Resonance Gas}}, \href{http://dx.doi.org/10.1103/PhysRevC.74.044903}{\emph{Phys. Rev.} {\bf C74} (2006) 044903}, [\href{http://arxiv.org/abs/nucl-th/0606036}{{\tt nucl-th/0606036}}].

\bibitem{qcd-phase-fluctuations-review-02}
X.~Luo and N.~Xu, \emph{{Search for the QCD Critical Point with Fluctuations of Conserved Quantities in Relativistic Heavy-Ion Collisions at RHIC : An Overview}}, \href{http://dx.doi.org/10.1007/s41365-017-0257-0}{\emph{Nucl. Sci. Tech.} {\bf 28} (2017) 112}, [\href{http://arxiv.org/abs/1701.02105}{{\tt 1701.02105}}].

\bibitem{statistical-model-10}
P.~Garg, D.~K. Mishra, P.~K. Netrakanti, B.~Mohanty, A.~K. Mohanty, B.~K. Singh et~al., \emph{{Conserved number fluctuations in a hadron resonance gas model}}, \href{http://dx.doi.org/10.1016/j.physletb.2013.09.019}{\emph{Phys. Lett.} {\bf B726} (2013) 691--696}, [\href{http://arxiv.org/abs/1304.7133}{{\tt 1304.7133}}].

\bibitem{statistical-model-08}
J.-H. Fu, \emph{{Higher moments of multiplicity fluctuations in a hadron-resonance gas with exact conservation laws}}, \href{http://dx.doi.org/10.1103/PhysRevC.96.034905}{\emph{Phys. Rev.} {\bf C96} (2017) 034905}, [\href{http://arxiv.org/abs/1610.07138}{{\tt 1610.07138}}].

\bibitem{urqmd-10}
M.~Bleicher et~al., \emph{{Relativistic hadron hadron collisions in the ultrarelativistic quantum molecular dynamics model}}, \href{http://dx.doi.org/10.1088/0954-3899/25/9/308}{\emph{J. Phys. G} {\bf 25} (1999) 1859--1896}, [\href{http://arxiv.org/abs/hep-ph/9909407}{{\tt hep-ph/9909407}}].

\bibitem{hydro-fluctuations-02}
L.~Jiang, P.~Li and H.~Song, \emph{{Correlated fluctuations near the QCD critical point}}, \href{http://dx.doi.org/10.1103/PhysRevC.94.024918}{\emph{Phys. Rev.} {\bf C94} (2016) 024918}, [\href{http://arxiv.org/abs/1512.06164}{{\tt 1512.06164}}].

\bibitem{sph-bes-01}
H.-H. Ma, D.~Wen, K.~Lin, W.-L. Qian, B.~Wang, Y.~Hama et~al., \emph{{Hydrodynamic results on multiplicity fluctuations in heavy-ion collisions}}, \href{http://dx.doi.org/10.1103/PhysRevC.101.024904}{\emph{Phys. Rev.} {\bf C101} (2020) 024904}, [\href{http://arxiv.org/abs/1910.00705}{{\tt 1910.00705}}].

\bibitem{RHIC-star-bes-11}
{\scshape STAR} collaboration, T.~J. Tarnowsky, \emph{{Charge Dependence and Scaling Properties of Dynamical $K/\pi$, $p/\pi$, and $K/p$ Fluctuations from the STAR Experiment}}, \href{http://dx.doi.org/10.5506/APhysPolBSupp.5.515}{\emph{Acta Phys. Polon. Supp.} {\bf 5} (2012) 515--522}, [\href{http://arxiv.org/abs/1201.3336}{{\tt 1201.3336}}].

\bibitem{RHIC-star-bes-12}
{\scshape STAR} collaboration, J.~Thader, \emph{{Higher Moments of Net-Particle Multiplicity Distributions}}, \href{http://dx.doi.org/10.1016/j.nuclphysa.2016.02.047}{\emph{Nucl. Phys.} {\bf A956} (2016) 320--323}, [\href{http://arxiv.org/abs/1601.00951}{{\tt 1601.00951}}].

\bibitem{RHIC-star-bes-20}
{\scshape STAR} collaboration, J.~Adam et~al., \emph{{Nonmonotonic Energy Dependence of Net-Proton Number Fluctuations}}, \href{http://dx.doi.org/10.1103/PhysRevLett.126.092301}{\emph{Phys. Rev. Lett.} {\bf 126} (2021) 092301}, [\href{http://arxiv.org/abs/2001.02852}{{\tt 2001.02852}}].

\bibitem{RHIC-star-bes-21}
{\scshape STAR} collaboration, M.~S. Abdallah et~al., \emph{{Measurements of Proton High Order Cumulants in $\sqrt{s_{_{\mathrm{NN}}}}$ = 3 GeV Au+Au Collisions and Implications for the QCD Critical Point}}, \href{http://dx.doi.org/10.1103/PhysRevLett.128.202303}{\emph{Phys. Rev. Lett.} {\bf 128} (2022) 202303}, [\href{http://arxiv.org/abs/2112.00240}{{\tt 2112.00240}}].

\bibitem{RHIC-star-bes-22}
{\scshape STAR} collaboration, J.~Adam et~al., \emph{{Flow and interferometry results from Au+Au collisions at $\sqrt{s_{NN}} = 4.5$ GeV}}, \href{http://dx.doi.org/10.1103/PhysRevC.103.034908}{\emph{Phys. Rev. C} {\bf 103} (2021) 034908}, [\href{http://arxiv.org/abs/2007.14005}{{\tt 2007.14005}}].

\bibitem{RHIC-star-bes-25}
{\scshape STAR} collaboration, Z.~Sweger, \emph{{Recent Results and Future Prospects from the STAR Beam Energy Scan Program}},  in \emph{{57th Rencontres de Moriond on QCD and High Energy Interactions}}, 5, 2023.
\newblock \href{http://arxiv.org/abs/2305.07139}{{\tt 2305.07139}}.

\bibitem{Cardoso:2004nk}
V.~Cardoso, O.~J.~C. Dias, J.~P.~S. Lemos and S.~Yoshida, \emph{{The Black hole bomb and superradiant instabilities}}, \href{http://dx.doi.org/10.1103/PhysRevD.70.049903}{\emph{Phys. Rev. D} {\bf 70} (2004) 044039}, [\href{http://arxiv.org/abs/hep-th/0404096}{{\tt hep-th/0404096}}].

\bibitem{Brito:2015oca}
R.~Brito, V.~Cardoso and P.~Pani, \emph{{Superradiance}: {New Frontiers in Black Hole Physics}}, \href{http://dx.doi.org/10.1007/978-3-319-19000-6}{\emph{Lect. Notes Phys.} {\bf 906} (2015) pp.1--237}, [\href{http://arxiv.org/abs/1501.06570}{{\tt 1501.06570}}].

\end{thebibliography}\endgroup
\end{document}